\documentclass[cernpreprint,11pt,a4paper,atlasdraft=false, texlive=2023, UKenglish, texmf, orcidlogo]{atlasdoc}
 
\usepackage{atlaspackage}
\usepackage{atlasbiblatex}
 
\usepackage{atlasphysics}
 
\graphicspath{{logos/}{figures/}}
 
\usepackage{subfig}
 
\usepackage{ulem}
\usepackage{siunitx}

\AtlasTitle{Measurement of exclusive pion pair production in proton--proton collisions at $\sqrt{s} = 7$ TeV with the ATLAS detector}

\AtlasJournalRef{Eur. Phys. J. C 83 (2023) 627}
\AtlasDOI{10.1140/epjc/s10052-023-11700-x}
 
\AtlasAbstract{
The exclusive production of pion pairs in the process $pp\to pp\pi^+\pi^-$  has been measured at $\sqrt{s} = 7$ TeV with the ATLAS detector at the LHC, using $\SI{80}{\micro\mathrm{b}}^{-1}$ of low-luminosity data. The pion pairs were detected in the ATLAS central detector while outgoing protons were measured in the forward ATLAS ALFA detector system. This represents the first use of  proton tagging to measure an exclusive hadronic final state at the LHC.  A cross-section measurement is performed in two kinematic regions defined by the proton momenta, the pion rapidities and transverse momenta, and the pion--pion invariant mass.
Cross-section values of \(4.8 \pm 1.0 \mathrm{\ (stat)} \uncertupdown{0.3}{0.2} \mathrm{\ (syst)}\ \SI{}{\micro\mathrm{b}}\) and \(9 \pm 6 \mathrm{\ (stat)} \uncertupdown{2}{2} \mathrm{\ (syst)}\ \SI{}{\micro\mathrm{b}} \) are obtained in the two regions; they are compared with  theoretical models and provide a demonstration of the feasibility of measurements of this type.
}
 
\author{The ATLAS Collaboration}
 
\AtlasRefCode{STDM-2017-07}

\AtlasDate{\today}
 
\PreprintIdNumber{CERN-EP-2022-140}

\hypersetup{pdftitle={ATLAS document},pdfauthor={The ATLAS Collaboration}}
 
\newcommand*{\betastar}{\ensuremath{\beta^{\star}}}

\renewcommand{\arraystretch}{1.2}
\newcommand{\as}[1]{\renewcommand{\arraystretch}{#1}}
\newcommand{\centercell}[1]{\multicolumn{1}{c}{#1}}
\DeclareExpandableDocumentCommand{\multihead}{m m}{\multicolumn{#1}{c}{\bfseries #2}}
\DeclareExpandableDocumentCommand{\head}{m}{\centercell{\bfseries #1}}
\DeclareExpandableDocumentCommand{\SIhead}{m}{\multicolumn{1}{s}{\bfseries #1}}
\DeclareSIUnit\murad{\micro\radian}
\newcommand*{\GENEX}{\textsc{GenEx}\xspace}
\newcommand*{\PYTHIAe}{\textsc{Pythia\,8}\xspace}          
\newcommand*{\GEANTf}{\textsc{Geant4}\xspace}

\newcommand*{\DIME}{\textsc{Dime}\xspace}
\DeclareSIUnit\ub{{\micro}b}
\newcommand*{\ub}{\mbox{\si{\ub}}}
\newcommand*{\iub}{\mbox{\si{\ub$^{-1}$}}}
\newcommand{\SIspace}{\ensuremath{\mskip3mu}}
\newcommand*{\uncertupdown}[2]{{~}^{+#1}_{-#2}\xspace}
 
\DeclareUnicodeCharacter{2212}{-}
\renewbibmacro*{url+urldate}{}
 
\usepackage{makecell}
 
\usepackage{multirow}
 
\usepackage[hypcap=false]{caption}

\begin{document}
 
\maketitle
 
\tableofcontents
 
\section{Introduction}
A significant fraction of LHC proton--proton (\(pp\)) collisions are either elastic (\(\approx \SI{25}{\%}\)) or diffractive (\(\approx \SI{20}{\%}\) of the remaining inelastic collisions \cite{Zyla:2020zbs}). In many of these collisions the momentum transferred from the beam protons is too small (\(< 1\ {\GeV} \)) for perturbative quantum chromodynamics (QCD) to be applicable, and instead Regge theory \cite{Eden_1971,Jacob:1985bc} is used. This involves the exchange of a non-perturbative Pomeron, a soft colour-singlet object originally introduced to keep the universality of the \(s\)-dependence of high energy total cross sections by Gribov \cite{gribov1961possible}, Chew and Frautschi \cite{PhysRevLett.7.394} and later providing an explanation of the rise of hadronic collision cross-sections with increasing beam energy energy~ \cite{Nachtmann_2004}.  Since QCD is assumed to be universally valid, there should nevertheless be a `mapping' between Regge theory and QCD in which the exchanged Pomeron
is equivalent to the exchange of quarks and gluons, the fundamental quanta of QCD.
In certain kinematic regions, the Pomeron can be treated approximately as a two-gluon colour-singlet state. Exclusive diffractive processes provide a clean environment to study the interactions of Pomerons, via the observation of final states involving only several hadrons and no beam-remnants except the outgoing protons.
 
\begin{figure}[ht!]
\centering
\includegraphics[width=0.8\textwidth]{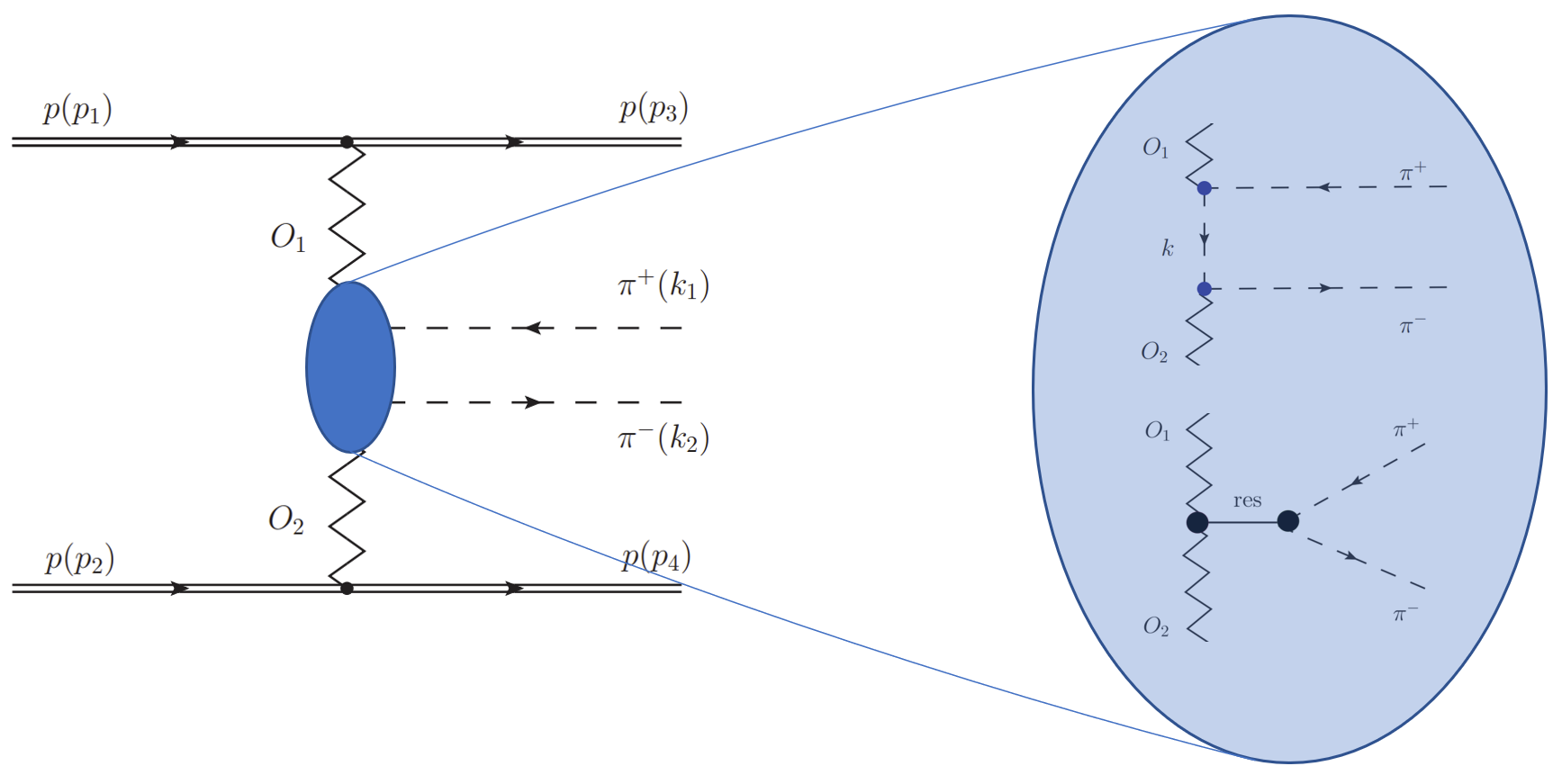}
\caption{A schematic Feynman diagram (left) describing the process of interest, in which two protons \(p\) with momenta \(p_1\), \(p_2\)  interact to give a pion pair \(\pi^{+}\pi^{-}\) with momenta \(k_1\), \(k_2\) accompanied by two outgoing protons with momenta \(p_3\), \(p_4\). The exchanged virtual objects \(O_1\) and \(O_2\) are not necessarily of the same type, but may be typified as Reggeons (e.g. the Pomeron) and must have all hadronic quantum numbers equal to zero. Two of the simplest examples of possible production mechanisms are shown in the bubble (right) -- the upper one belonging to so-called continuum contributions and the lower one to resonance contributions.}
\label{fig:pipi_schematics}
\end{figure}

This paper presents first measurements of the exclusive double-diffractive production of charged-pion pairs in the process $pp\to pp\pi^+\pi^-$, schematically shown in \Fig{\ref{fig:pipi_schematics}}, using the ATLAS detector at the LHC.
 
The charged pions are measured with the ATLAS inner detector~\cite{PERF-2007-01}, while the outgoing protons are measured by the ALFA detector \cite{1748-0221-11-11-P11013}, comprising two tracking stations situated in the forward regions on either side of ATLAS at distances of  approximately \SI{240}{m} from the ATLAS interaction point (IP). The ALFA detector system was designed in the first place to determine the total $pp$ cross section at LHC energies, as well as to measure the LHC luminosity via small-angle proton-scattering processes \cite{Aad:2014dca}.
The use of the ALFA detectors is a distinctive feature of the present study, since the outgoing protons are directly measured, suppressing contamination from the remnants of excited scattered protons. The analysis benefits from procedures developed previously in the ALFA total cross section measurement \cite{Aad:2014dca}.
 
The present results can be compared with existing measurements of the exclusive production of pion pairs using data from the ISR \cite{Breakstone1989, Breakstone:1990at}, SPS \cite{Antinori:1995wz}, RHIC \cite{star:2020excl} and LHC \cite{CMS-FSQ-12-004, Sirunyan2020}, where the CMS results \cite{CMS-FSQ-12-004, Sirunyan2020} are not based on purely exclusive events but on a mixture of exclusive and semi-exclusive processes.
Some recent theory predictions for the process considered here may be found in Refs.~\cite{Ryutin_2019, Pelaez_2019, Godizov_2016, Lebiedowicz:2016ioh, LebPhysRevD.91.074023, Bolz_2015, Ryutin:2014eua, Lebiedowicz:2014t, Lebiedowicz:2009pj, Harland-Lang:2013dia, Albrow:2010yb}.
In addition to revealing general properties of Pomerons, these channels provide a rich potential harvest for a variety of final states, including the production of resonances and the possible production of the long-sought glueball states.
 
\section{Experimental apparatus}
 
ATLAS is a multipurpose apparatus covering almost the entire solid angle around its LHC interaction point (IP)~\cite{PERF-2007-01}.\footnote{ATLAS uses a right-handed coordinate system with its origin at the nominal interaction point in the centre of the detector and the $z$-axis along the beam-pipe. The $x$-axis points from the interaction point to the centre of the LHC ring, and the $y$-axis points upwards. Cylindrical coordinates $(r,\phi)$ are used in the transverse plane, $\phi$ being the azimuthal angle around the $z$-axis. The pseudorapidity is defined in terms of the polar angle $\theta$ as $\eta = -\ln \tan(\theta/2)$. Angular distance is measured in units of $\Delta R \equiv \sqrt{(\Delta\eta)^{2} + (\Delta\phi)^{2}}$.}
The present analysis  combines information from the ATLAS inner detector, the minimum-bias trigger scintillators, and the ALFA detector system.
 
The ATLAS inner detector (ID) \cite{PERF-2007-01, ATLAS:1997ag}
immediately surrounds the ATLAS IP and is composed of three tracker subsystems that provide high-precision track reconstruction: a~silicon pixel detector (innermost), a~silicon semiconductor tracker (SCT), and a transition-radiation tracker (outermost), which also helps to discriminate electrons from hadrons. The ID covers a range of \(|\eta|<2.5\). It is surrounded by a superconducting solenoid, which produces a~\SI{2}{T} axial field within the ID volume.

The minimum-bias trigger scintillators (MBTS) \cite{Kwee:1140965} are located on either side of ATLAS  between the ID and the liquid-argon calorimeter at  \(\pm3.6~\textrm{m}\) from the IP. Each of the two MBTS detectors is oriented perpendicularly to the beam direction and consists of 16 panels assembled in two concentric rings, the outer ring covering the range \(2.1<|\eta|<2.8\) and the inner ring  covering the range \(2.8<|\eta|<3.8\).
 
\begin{figure}[ht!]
\centering
\includegraphics[width=0.8\textwidth]{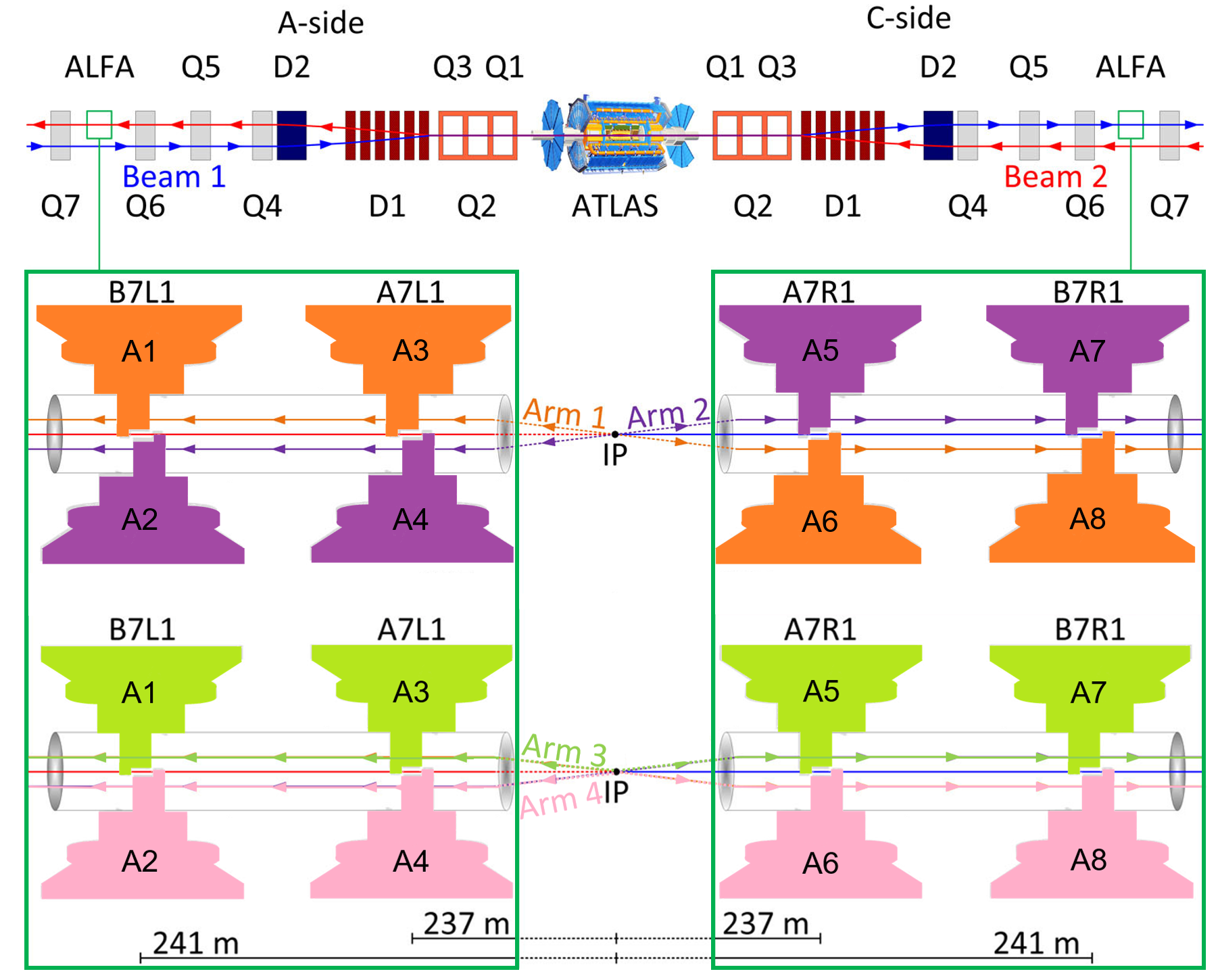}
\caption{A schematic depiction of the arrangement of ALFA in ATLAS, indicating the naming and numbering conventions used \cite{1748-0221-11-11-P11013}. The upper part of the diagram gives a plan view that shows the dipole (D) and quadrupole (Q) magnets on either side of ATLAS whose settings are relevant for the beam optics. The proton beams that circulate anticlockwise and clockwise are depicted in this view, emerging into the A and C sides, respectively. The lower diagrams give side views that show the  ALFA detector combinations (\textbf{arms}) used for the elastic measurement (Arm~1, orange, and  Arm~2, purple) and the anti-elastic measurement (Arm~3, green, and Arm~4, pink). Each arm comprises two sections (\textbf{armlets}), consisting of two upper or lower detectors on a single side. The ALFA stations are numbered from A1 to A8 together with a more detailed designation. In these views the vertically deflected protons entering the different arms are indicated.}
\label{fig:alfa_scheme}
\end{figure}

\begin{figure}[ht!]
\centering
\includegraphics[width=0.6\textwidth]{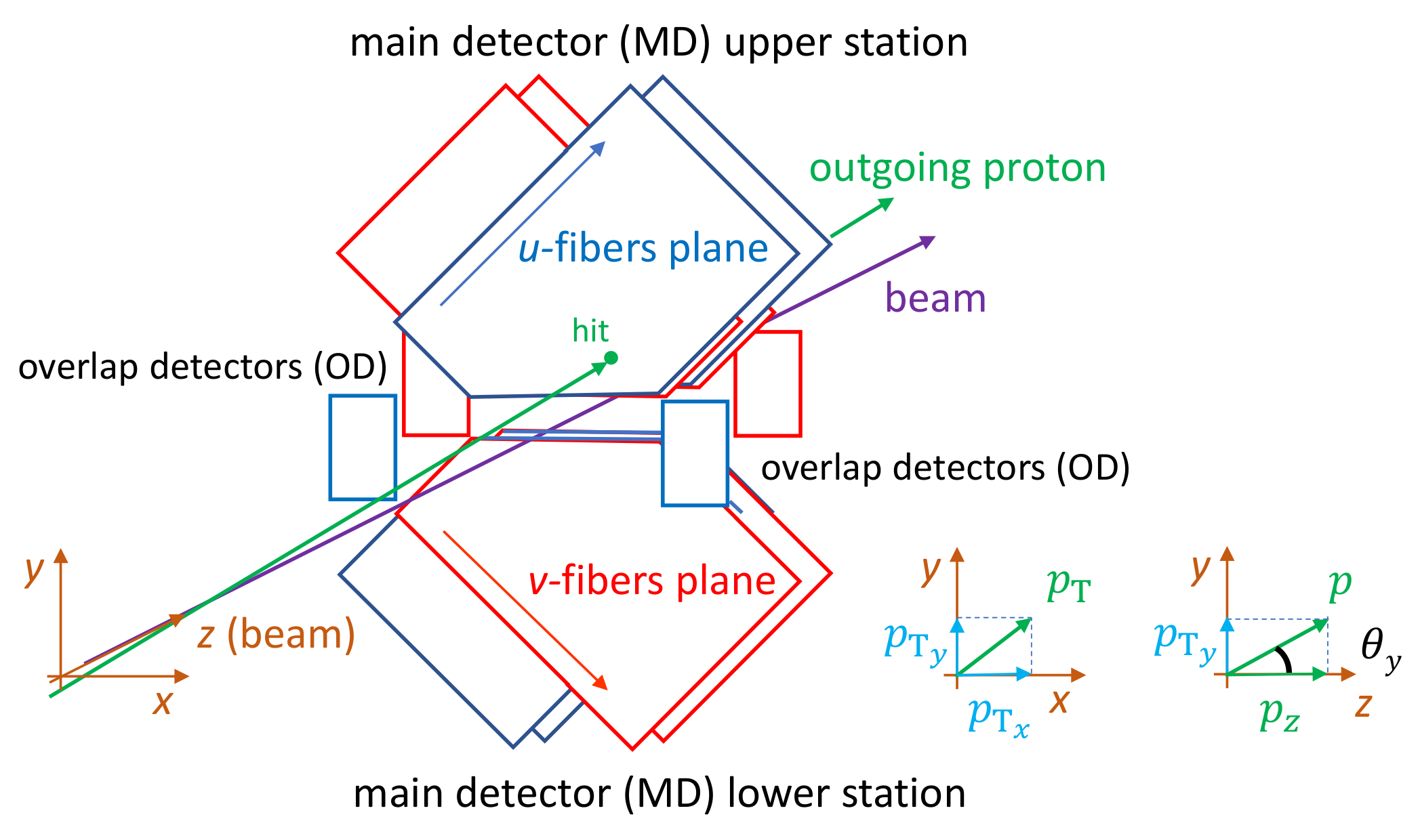}
\caption{A schematic drawing of the ALFA detectors in one station, here A5+A6 in  \Fig{\ref{fig:alfa_scheme}}. Each of the main detectors is made of ten \(u\) and ten \(v\) fibre layers; for more details see Ref.~\cite{1748-0221-11-11-P11013}. Only the first and last layers are drawn. For the overlap detectors, serving for the mutual positioning of the upper and lower detectors, only one of the three planes is drawn. The trigger planes are not drawn. The local coordinate system, relative to the beamline, is indicated  together with the projections of the leading outgoing proton momentum \(p\) and the angle \(\theta_y\); the angle \(\theta_x\) is defined similarly. }
\label{fig:ALFA_detector_schematics}
\end{figure}

The ALFA detector system \cite{1748-0221-11-11-P11013}, located at $z \approx ~\pm\SI{240}{m}$, consists of four Roman pot stations, containing vertically movable detectors that can approach the outgoing beams to within one millimetre. A schematic diagram of the detector arrangement relative to the ATLAS IP is shown in \Fig{\ref{fig:alfa_scheme}}.
Each station is equipped with two Roman pots, each of which hosts a nearly edgeless tracking detector composed of scintillating fibres, as depicted in  \Fig{\ref{fig:ALFA_detector_schematics}}.
The ALFA detector system has a limited radiation tolerance and takes data under conditions with low numbers (\(\mu\)) of simultaneous proton--proton collisions interactions per beam crossing (pile-up).
 
An extensive software suite~\cite{ATL-SOFT-PUB-2021-001} is used in the reconstruction and analysis of real and simulated data, in detector operations, and in the trigger and data acquisition systems of the experiment.
 
\section{Monte Carlo generators}
 
The analysis makes use of two Monte Carlo (MC) signal generators: \GENEX \cite{Kycia:2014hea} and \DIME \cite{Harland-Lang:2013dia}.\footnote{Some other generators, also suitable for  modelling the considered process, became available later, e.g.\ ExDiff \cite{ryutin2018exdiff} and SuperChic~\cite{Harland_Lang_2016}.}
The models are similar, but not identical, and provide different predictions. The \GENEX\ generator is used for the baseline calculations of detection and reconstruction efficiency for the events,  and for corrections to the data, while \DIME\ is used for comparison and to estimate model uncertainties.
 
The generator \GENEX\ (version May 2015) is based on a theoretical model from Lebiedowicz et al.~\cite{Lebiedowicz:2009pj} and calculates the exclusive continuum production of \(\pi^{+}\pi^{-}\) and \(K^{+}K^{-}\) pairs. It is based on a Regge-exchange description. The parameter determining the meson form factor (in its exponential parameterization) is the only free parameter. The implemented model describes non-resonant production without an absorption correction and does not include a rapidity gap survival factor (see e.g.{} Ref.~\cite{Khoze:2017sdd}).
In the present analysis, pions were generated within a region of pseudorapidity \(|\eta|<2.7\), and with the off-shell-pion form-factor parameter set to \(\Lambda_{\mathrm{off}} = \SI{1}{\GeV}\).  Otherwise, default program settings were used.
 
The generator \DIME\ (version 1.06) similarly generates exclusive continuum production of \(\pi^{+}\pi^{-}\) and \(K^{+}K^{-}\) pairs, although other channels are also implemented, e.g.\ exclusive \(\rho\rho\) or \(\phi\phi\) production. Unlike  \GENEX, \DIME\ includes wider possibilities for modelling absorption effects: four different models (parameter \(iin\), 1 - the default value used in our analysis; see Table 12 in Ref.~\cite{Khoze_2013}) for absorption are available with three different parameterizations of the meson form factor (exponential, `Orear-like' -- used in the present analysis, and power-like)~\cite{Harland-Lang:2013dia}.
 
The central diffractive background processes were generated by \PYTHIAe~\cite{Sjostrand_2008}
(version 8.183), using the \mbox{ATLAS A2} set of tuned parameters \cite{ATL-PHYS-PUB-2011-014} and the \textsc{MSTW2008lo} PDF set \cite{Watt_2012}.  \PYTHIAe employs a parton structure function approach and calculates the processes of central diffraction via double Pomeron exchange (DPE). The exclusive pion-pair process is removed from the dataset.
 
The \GEANTf~\cite{Agostinelli:2002hh} toolkit is implemented to simulate the response of the ATLAS central and forward detectors~\cite{SOFT-2010-01}. The leading outgoing protons are transported to the ALFA locations using the beam parameter values and ALFA detector positions corresponding to the data-taking period \cite{Aad:2014dca}
in which a special high-beta* optics, referred to  here as the nominal optics, was employed.  The MC events were reconstructed using the same ATLAS software chain as used for the data.
 
\section{Data taking}
\label{sec:data_taking}
The data were recorded in October 2011 during a special four-hour run with high \(\betastar = \SI{90}{m}\) and very low pile-up given by $\mu = 0.035$. Each beam of the LHC was filled with one circulating proton bunch containing \(7\cdot10^{10}\) protons.
The integrated delivered luminosity and its uncertainty were determined to be
\(L_{\textrm{int}} = 78.7 \pm 0.1 {(\textrm{stat})} \pm 1.9 {\textrm{ (syst)}} \SIspace \iub\)~\cite{Aad:2014dca}.
 
Two event topological configurations are used in this analysis, covering four regions of phase space. They are referred to as  \textbf{elastic} and \textbf{anti-elastic} and are  based on signals in different  combinations of four ALFA detectors, as shown in Figure~\ref{fig:alfa_scheme}. Each detector combination is designated as an \textbf{arm} (following the existing ALFA notation \cite{Aad:2014dca}), and includes detectors on side A and side C in order to measure forward protons on each side of ATLAS.
The elastic configuration  makes use of  two arm combinations: Arm1 = arm\(_{1368}\) = {\small A1+A3+A6+A8} (\textbf{elastic arm 0}) and Arm2 = arm\(_{2457}\) = {\small A2+A4+A5+A7} (\textbf{elastic arm 1}), while the anti-elastic configuration uses Arm3 = arm\(_{1357}\) = {\small A1+A3+A5+A7} (\textbf{anti-elastic arm 0}) and Arm4 = arm\(_{2468}\) = {\small A2+A4+A6+A8} (\textbf{anti-elastic arm 1}). Each arm comprises two \textbf{armlets}, consisting of two ALFA stations on either the A side or C side of ATLAS.  The two configurations measure different ranges of momentum transfer to the pion system and their events were recorded by different combinations of triggers during data taking.  They correspond physically to different kinematic phase-space regions, and theoretical models can differ in their estimates of cross-sections.
 
\section{Data analysis}
 
\subsection{Event selection}
\label{sec:data_analysis_event_selection}
 
The event topology for exclusive pion-pair production consists of two oppositely charged pions detected by the ATLAS ID, accompanied by a leading outgoing proton entering each of the ATLAS forward regions and detected by the ALFA detector system.
There should be no other activity in the ATLAS ID and in the inner cells of the MBTS, and the total three-momentum of the measured protons and pions must be consistent with zero.\footnote{The precise expectation value depends on the beam-crossing angle, which varied during the data taking and was at most \(\SI{10}{\murad}\) in the  $y$ projection and smaller in $x$. The full effect is safely ignored since its impact is less than those of the ALFA detector resolution.} The data selections are summarized in Table~\ref{tab:cuts}. The order in which the selections were applied to the data is given in \Tab{\ref{tab:data_statistics}}.

\begin{minipage}{0.95\textwidth}
\centering
 
\captionof{table}{List of applied selections.}
\label{tab:cuts}
\begin{tabular}[l]{@{}l@{}}
\toprule{Selection} \\ [0.5ex]
\midrule
Bunch selection   \\
Lumi blocks selection  \\
Trigger configuration  \\
Pions: \\
 
\ \  \ \ number of tracks \\
\ \ \ \ primary vertex \\
\ \ \ \ ID track quality  \\
MBTS veto \\
Protons: \\
\ \ \ \ ALFA track quality \\
\ \ \ \ ALFA \textit{uv}-condition \\
\ \ \ \ ALFA clean track \\
\ \ \ \ ALFA geometry condition  \\
Full system momentum balance in $p_{x}$ and $p_{y}$ \\
Fiducial region \\
\bottomrule
\end{tabular}
\end{minipage}

\subsubsection{Description of selections}
\label{sec:data_analysis_event_selection_description}
 
The description of the conditions that are applied to select a clean sample of $pp\pi^+\pi^-$ event candidates, as summarized in Table~\ref{tab:cuts}, is as follows.
\begin{itemize}
\item{\textbf{Bunch system and good lumi blocks.} The data come from single colliding proton bunches in the LHC. The data-recording intervals that were used (`lumi blocks') are required to be longer than 60~seconds in duration and the dead time must be smaller than 5\%.}
\item{\textbf{Trigger.} For the elastic configuration, a coincidence in the upper A-side and lower C-side detectors or the lower A-side and upper C-side detectors was required \cite{Aad:2014dca} (see \Fig{\ref{fig:alfa_scheme}}). The trigger efficiency within the geometrical acceptance of the detectors is measured to be \((99.96 \pm 0.10)\%\) \cite{Aad:2014dca}.
For the anti-elastic configuration the relevant trigger (a signal in any of the eight ALFA detectors) was prescaled by a factor of 15, which resulted in a low numbers of recorded anti-elastic events. The trigger efficiency for each of the anti-elastic configurations is determined as the product of  the measured single-diffractive trigger efficiencies on each side, leading to the value \((99.02 \pm 0.25)\%\)  before prescaling.}
\item{\textbf{Pion system.}
\begin{itemize}
\item{\textbf{Charged tracks.}  Tracks in the ATLAS ID were reconstructed using a low-$p_{\mathrm{T}}$ tracking algorithm (\(p_{\mathrm{T}} > \SI{100}{\MeV}\)) as in Ref.~\cite{STDM-2010-06}, where \(p_{\mathrm{T}}\) denotes the transverse momentum component. Each track must  pass the following quality conditions: 
\begin{description}
\item[---] \textbf{Pixel detector:} at least one hit is required,
\item[---] \textbf{Semiconductor tracker:} a minimum number of hits in the SCT is required, depending on the track $p_{\mathrm T}$: at least two hits for \(p_{\mathrm T} > \SI{100}{\MeV}\), four hits for \(p_{\mathrm T} > \SI{200}{\MeV}\) and six hits for \(p_{\mathrm T} > \SI{300}{\MeV}\).
\end{description}
Events with at least two tracks are required. A primary vertex is required and must have exactly two accepted tracks with opposite-sign charges and small impact parameters relative to the beam axis.  For the purposes of this analysis, all charged tracks are treated as pion tracks, since the numbers of electron, kaon and proton pairs are estimated to be small.
The pions' transverse momenta are reconstructed \cite{STDM-2010-06} with an uncertainty of about \SI{10}{\MeV} independent of their charge.
}

\end{itemize}
}
\item{\textbf{MBTS.} To select events with no additional visible activity in the ATLAS ID, taking into consideration its limited \(\eta\) coverage, a MBTS veto criterion is applied: the total number of MBTS inner cells  on the two ATLAS sides with a signal is required to be at most one, ignoring the outer cells. This criterion was chosen since it was robust against electronic noise in the MBTS detector.}
\item{\textbf{Proton system.}
\begin{itemize}
\item{\textbf{ALFA track reconstruction.} The proton reconstruction criteria used in a previous analysis~\cite{Aad:2014dca} -- bunch group, data quality, selections on the reconstructed track and geometrical deflection, beam-screen and edge conditions (see Table~2 of Ref.~\cite{Aad:2014dca}) -- together with the detector alignment, are adopted for the current study.}
 
\item{\textbf{ALFA \textit{uv}-condition.} This requirement is introduced to remove contributions from beam halo and showers (caused by interactions with material between the IP and the ALFA detectors). These contributions are absent in MC simulations. The purpose of the ALFA \textit{uv}-condition is to differentiate between events with single particles and those with multi-particle showers entering the detectors, so as to filter out multi-particle signals, and in this way to allow good correspondence between MC simulations and real data samples.
 
The \textit{uv}-condition is based on the fibre multiplicities associated with a candidate proton. The first  four {\textit{u}} and first four {\textit{v}} detector layers~\cite{1748-0221-11-11-P11013}, closest to the IP, must satisfy the following:
\begin{description}
\item[---] at least two of these $u$ layers, and at least two of these $v$ layers, must have at least one hit, and
\item[---] at most one {\textit{u}} layer and one {\textit{v}} layer may have more than three hits.
\end{description}
The first condition identifies a well-measured track and is based on the estimate of the single-fibre efficiency and a requirement on the {\textit{uv}}-condition efficiency. The second, allowing at most two cross-talk hits surrounding that of the signal, is based on observations made during beam tests of the ALFA detectors. It also rejects tracks that give rise to showers.}
 
\item{\textbf{ALFA clean track.} Due to e.g.\ beam halo, pile-up or showers, more than one track may be reconstructed in a given ALFA detector~\cite{Aad:2014dca}, and it may not be possible to decide which of the tracks is the correct one. Therefore, only events with exactly one reconstructed track in each of the detectors that comprise an arm are considered in the present analysis, and a correction is evaluated for each of the configuration arms to compensate for the consequent loss of signal.}
 
\item{\textbf{ALFA geometric condition.} In measuring leading protons, the ALFA detector system operates as a spectrometer in which a proton having a larger energy loss undergoes a larger deflection, corresponding to a larger  \(x\) value of its track position measured in the horizontal direction in the coordinates of the ALFA station.  In the case of exclusive pion-pair production, the energy loss of a leading proton is low and the expected distribution in the scatter plot of the reconstructed \(x\)-positions of the protons and corresponding local angles \(\theta_x\) (see \Fig{\ref{fig:ALFA_detector_schematics}}) should have a shape similar to that of elastically scattered protons. This is illustrated in  \Fig{\ref{fig:alfa_geom_corr_ellipse}}  and allows a geometrical selection to be used to remove many protons not originating from the  process of interest.  Data and MC events were selected within the ellipse indicated in \Fig{\ref{fig:alfa_geom_corr_ellipse_genex}}, whose size is slightly larger than that from the elastically scattered protons shown in \Fig{\ref{fig:alfa_geom_corr_ellipse_elastic}}.  This covers the larger kinematic phase space occupied by the protons after losing energy to the pion system.
\begin{figure}[ht!]
\centering
\subfloat[]{\includegraphics[width=0.5\textwidth]{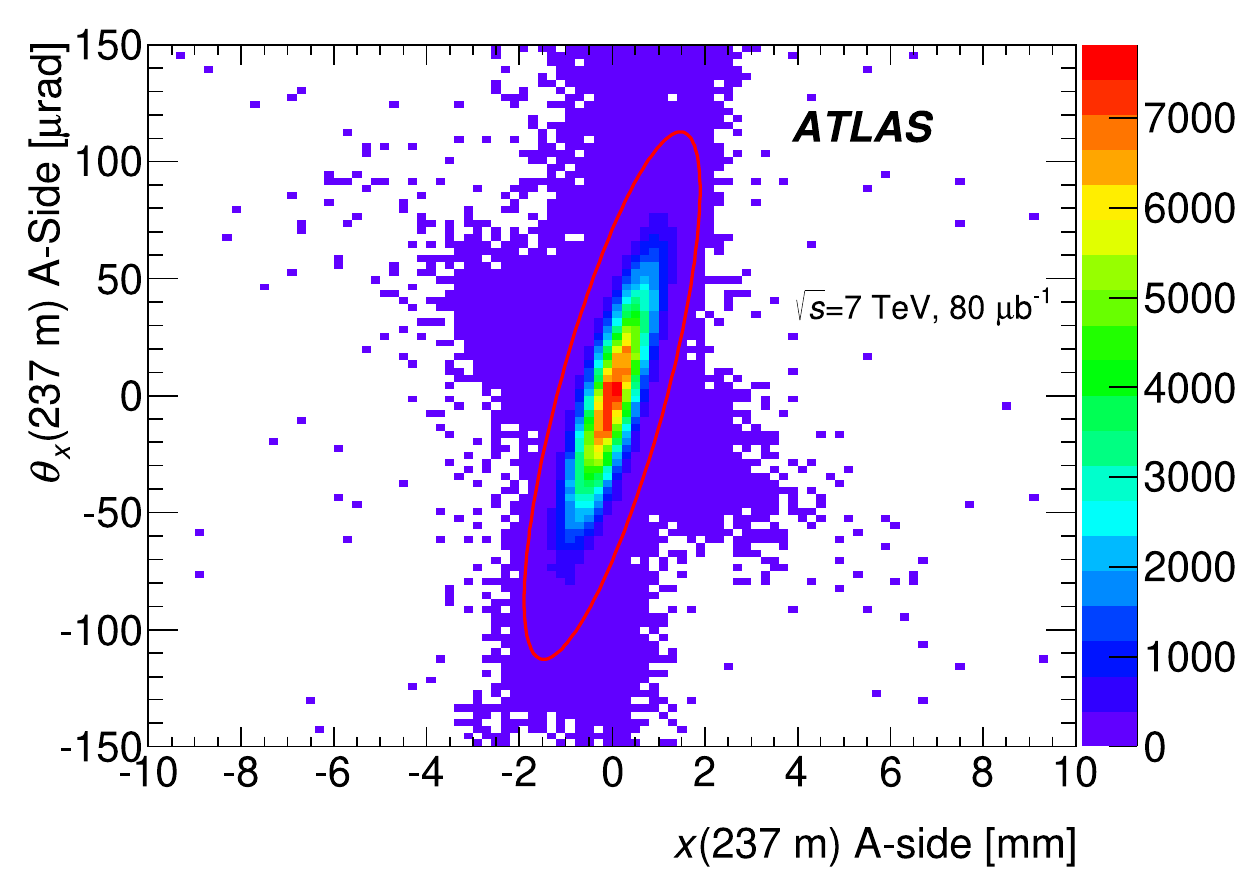}\label{fig:alfa_geom_corr_ellipse_elastic}}
\subfloat[]{\includegraphics[width=0.5\textwidth]{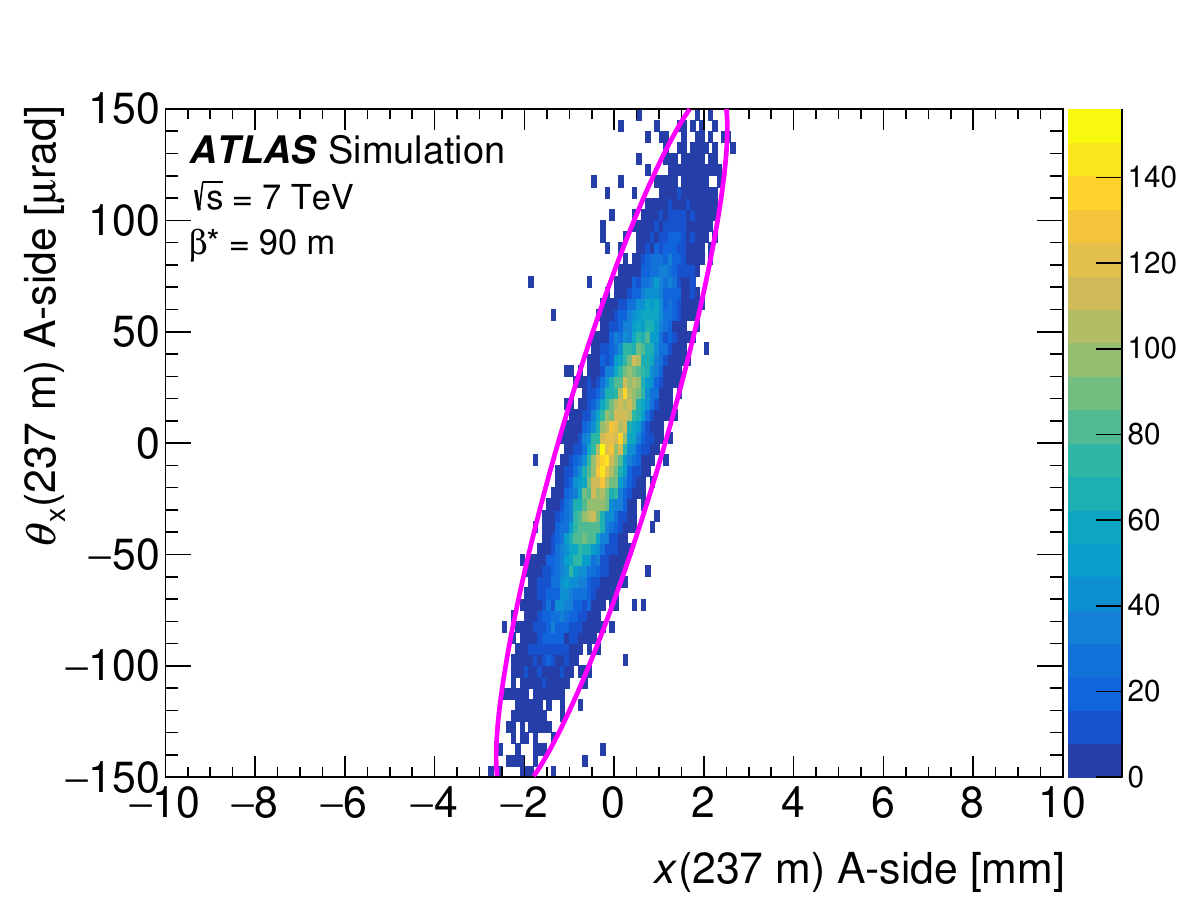}\label{fig:alfa_geom_corr_ellipse_genex}}
\caption{The correlation between the reconstructed position of a proton at $|z|=237$~m (A-side; a similar correlation is measured for the C-side)  and its local angle projection on the horizontal direction. In (a) the proton candidates in elastic proton--proton data lie within the area of the indicated ellipse \cite{Aad:2014dca}. The ellipse shown in (b) is obtained using the selection criteria applied to the sample of exclusive pion-pair production generated by \GENEX.}
\label{fig:alfa_geom_corr_ellipse}
\end{figure}}
\end{itemize}
}
 
\item{\textbf{Overall momentum balance.} Because of the limited reconstruction of the proton kinematics, a selection criterion for overall  four-momentum conservation is not possible; instead the momentum balance in the $x$ and $y$ projections is used.
 
The methods used for the reconstruction of the proton kinematics from the tracks measured by the ALFA detector are based on knowledge of the accelerator optics. As the mass of the centrally produced pion-pair system is low, the `lattice' method \cite{Aad:2014dca}, originally used for elastically scattered protons, can be applied. Proton momentum projections \(p_x\)  and \(p_y\) are used in the analysis, and  are reconstructed with a resolution of about \SI{40}{\MeV} and \SI{1}{\MeV} respectively.
 
The transverse momentum components of the full $pp\pi^+\pi^-$ system are determined with a resolution of approximately 60~\MeV\ in $p_x$ and 20~\MeV\ in $p_y$, denoted by $\sigma_x$ and $\sigma_y$ respectively.  The \textbf{signal region} is defined by $3.5\sigma_{x,y}$  intervals about zero in $p_x$ and  $p_y$.
}
 
\item{\textbf{Fiducial condition.}{}
The exclusive production of pion pairs is studied within a fiducial region which is defined to be close to that which is allowed by the sub-detector acceptances.
The data and Monte Carlo analyses use the same fiducial region, which is given in terms of the pseudorapidities \(\eta(\pi)\) and transverse momenta \(p_{\mathrm{T}}(\pi)\) of the individual pions, and the mass of the central pion system \(m_{\pi\pi}\):
\begin{equation*}
|\eta(\pi)| < 2.5, \; p_{\mathrm{T}}(\pi) > 0.1\ \GeV,\; 2m_{\pi} < m_{\pi\pi} < 2.0\ \GeV.
\end{equation*}
Additional conditions are applied to the $y$-components of the proton momenta \(p_{\mathrm{y}}(\mathrm{A})\) and \(p_{\mathrm{y}}(\mathrm{C})\), in both the  elastic and anti-elastic configurations:
 
\begin{equation*}
\mathrm{armlet}_{ij}^\mathrm{low\_cut} < p_{\mathrm{y}}(\mathrm{A}) < \mathrm{armlet}_{ij}^\mathrm{up\_cut} \ \mathrm{\ \ \ and \ \ \ }\  \mathrm{armlet}_{kl}^\mathrm{low\_cut} < p_{\mathrm{y}}(\mathrm{C}) < \mathrm{armlet}_{kl}^\mathrm{up\_cut},
\end{equation*}
 
where \(ijkl \in\) \{1368, 2457, 1357, 2468\} and

\begin{center}
\begin{math}
\begin{aligned}
\mathrm{armlet}_{13}^{\text{up\_cut}}\  &= \ \ 0.267\ \GeV,& \mathrm{armlet}_{57}^{\text{up\_cut}} &= \ \ 0.246\  \GeV,& \\
\mathrm{armlet}_{13}^{\text{low\_cut}}\  &= \ \ 0.080\  \GeV,& \mathrm{armlet}_{57}^{\text{low\_cut}} &= \ \ 0.080\  \GeV,&\\
\mathrm{armlet}_{24}^{\text{up\_cut}}\  &= -0.079\  \GeV,& \mathrm{armlet}_{68}^{\text{up\_cut}} &= -0.082\  \GeV,&\\
\mathrm{armlet}_{24}^{\text{low\_cut}}\  &= -0.250\  \GeV,& \mathrm{armlet}_{68}^{\text{low\_cut}} &= -0.269\  \GeV.&
\end{aligned}
\end{math}
\end{center}
 
These selections are evaluated in terms of the reconstructed ALFA proton kinematic parameters, calculated using the  nominal beam optics and ALFA detector positions. The experimental uncertainties of these values translate into a systematic uncertainty that is common to the corrected data and the MC predictions.}
 
\end{itemize}

\subsubsection{Data selection statistics}
The numbers of events after the successive selection conditions are shown in Table~\ref{tab:data_statistics}.
 
\begin{minipage}{0.95\textwidth}
\centering
\captionof{table}{Data statistics for the two configurations after the successive selection conditions.
Definitions of 'arm~0' and 'arm~1' (elastic and anti-elastic) are given in Section \ref{sec:data_taking}.}
\label{tab:data_statistics}
\as{1.2}
\begin{tabular}{@{}rrr@{}}
\toprule
\centercell{\bfseries Selection} & \multicolumn{2}{c}{\bfseries Configuration} \\
& elastic \ \ \ & {anti-elastic} \\
\midrule
Recorded ATLAS events & \multicolumn{2}{c}{6\,620\,953} \\
Data quality and trigger preselections & 1\,106\,855 & 397\,683 \\
ID selection (pion pair) & 1\,520 & 1\,115 \\
ALFA track selection (incl.\ clean track and \textit{uv}-condition)  & 486 & 11 \\
MBTS veto & 136 & 5 \\
ALFA geometry condition & 96 & 5 \\
Full system momentum balance in \(\mathit{x}\) and \(\mathit{y}\) & 30 & 3 \\
Fiducial region & 28 & 3 \\
\cline{1-3}
Total selected (arm0 + arm1) & 28 (18+10) & 3 (2+1) \\
\bottomrule
\end{tabular}
\end{minipage}
 
\subsection{Background estimation}
\begin{figure}[ht!]
\centering
\subfloat[]{\includegraphics[width=0.49\textwidth]{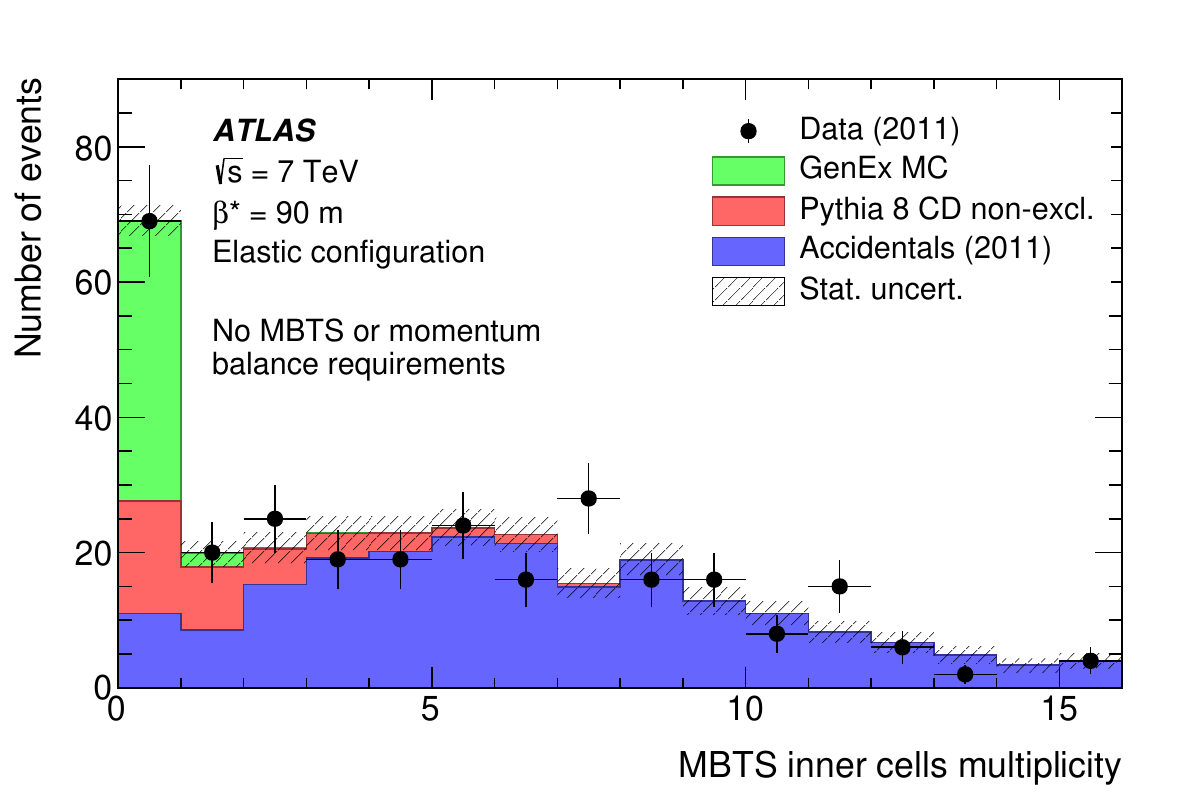}
}
\subfloat[]{\includegraphics[width=0.49\textwidth]{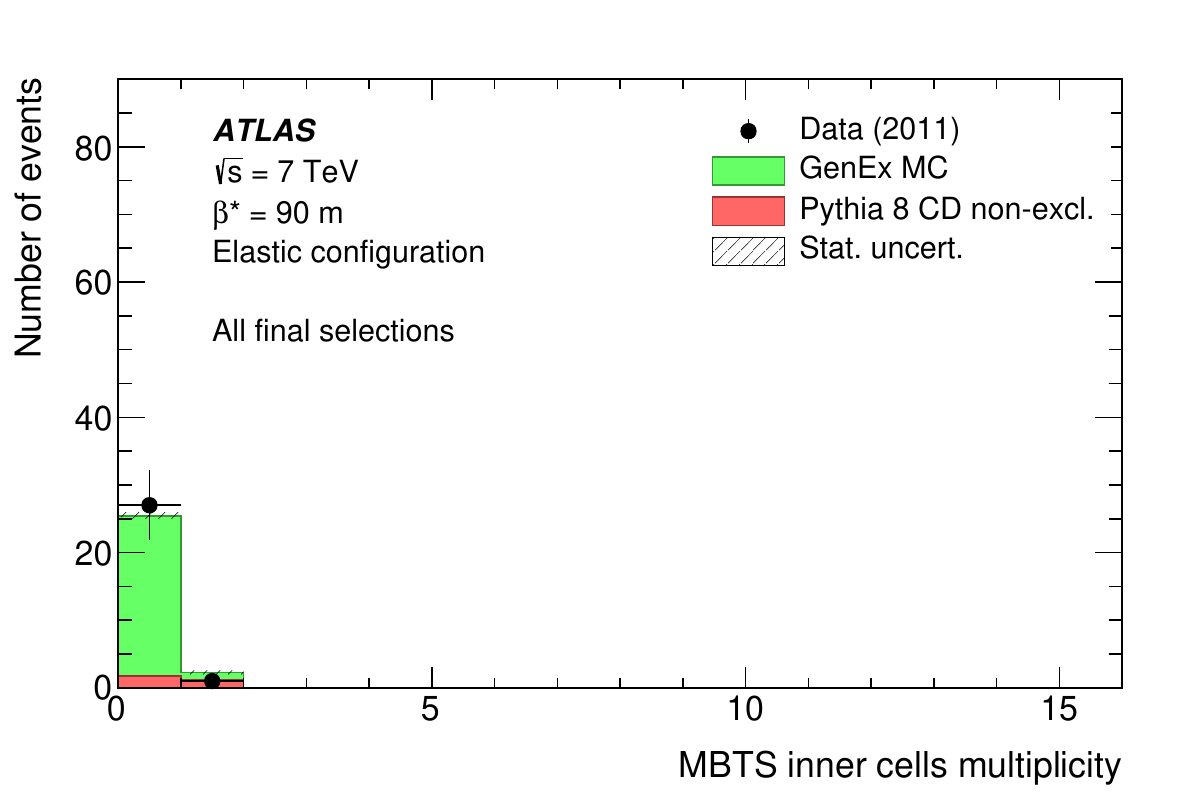}
\label{fig:background_accidentals_after_all_cuts}}
\caption{Total multiplicity of the MBTS inner cells, comparing data with  MC signal simulation (\GENEX), MC background simulation (\PYTHIAe central diffraction (CD)), and accidental coincidences. In (a) the MBTS selection criteria and selections for momentum balance in the $x$ and $y$ projections  are not applied; in (b) all the selections listed in \Tab{\ref{tab:cuts}} are applied. The MC and accidental coincidences histograms are stacked. Statistical uncertainties of MC events and accidentals are combined.}
\label{fig:background_accidentals}
\end{figure}

One source of background consists of centrally produced pion pairs that are accidentally accompanied by one or more beam-halo protons which are detected in ALFA.  This background is suppressed by the event selections to a low level.
This can be seen in \Fig{\ref{fig:background_accidentals}}, which shows (a) the total multiplicity distribution of the MBTS inner rings for events that satisfy the selection criteria
listed in Table~\ref{tab:cuts},
except for those on MBTS and momentum balance,
and (b) after all the selections in \Tab{\ref{tab:cuts}} are applied. A fit to the data is performed in terms of a model comprising generated signal events (\GENEX) and backgrounds from (i) generated diffractive events (\PYTHIAe) and (ii) an `accidentals' event sample formed from data events with a pion pair that are combined with unrelated data events randomly found to have ALFA signals.
For exclusive pion events, the MBTS multiplicity is at most one, while for other diffractive events it can be higher. Events with accidental coincidences can give both higher and  lower MBTS multiplicities. Such events are contained in the event sample, but \Fig{\ref{fig:background_accidentals_after_all_cuts}} shows that they contribute no contamination to the signal region $|p_a| < 3.5\sigma_a$, where $a = x, y$ (see Section~{\ref{sec:data_analysis_event_selection_description}}), 
after all conditions listed in \Tab{\ref{tab:cuts}} are applied.
 
\begin{figure}[ht!]
\centering
\subfloat[]{\includegraphics[width=0.49\textwidth]{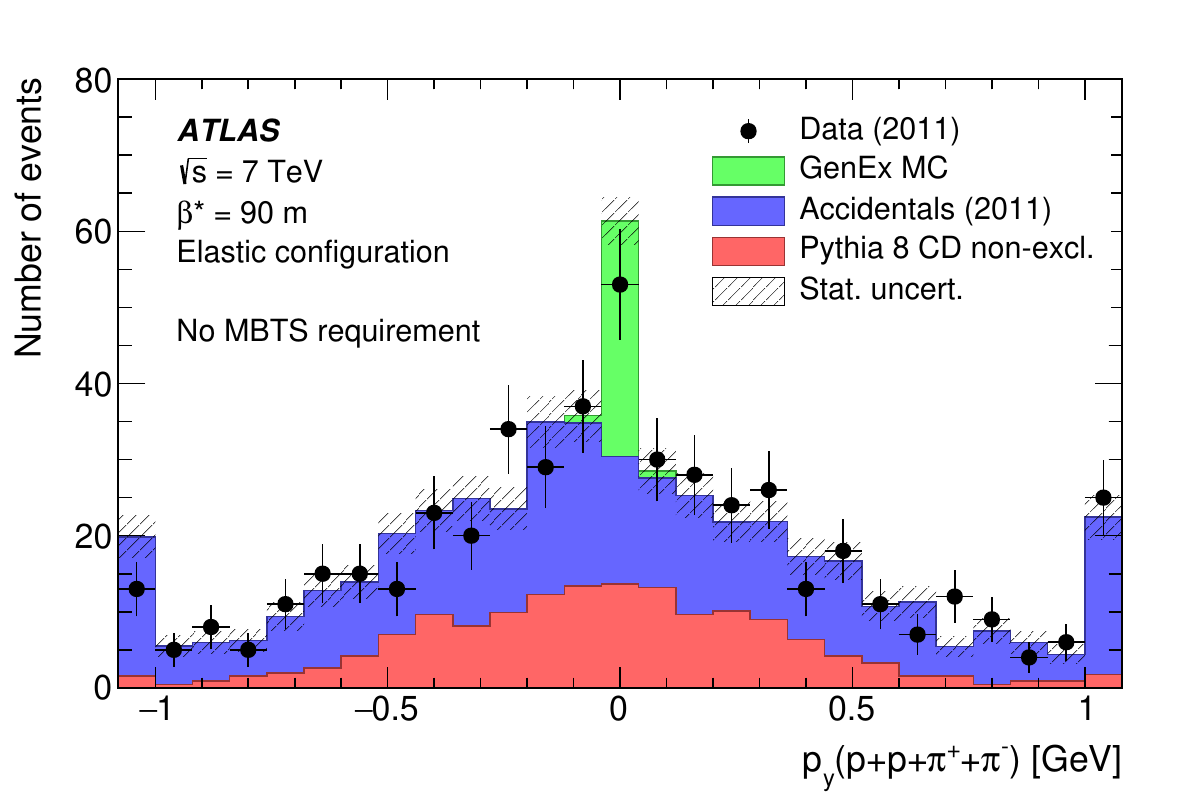}
\label{fig:background_before_MBTS}}
\subfloat[]{\includegraphics[width=0.49\textwidth]{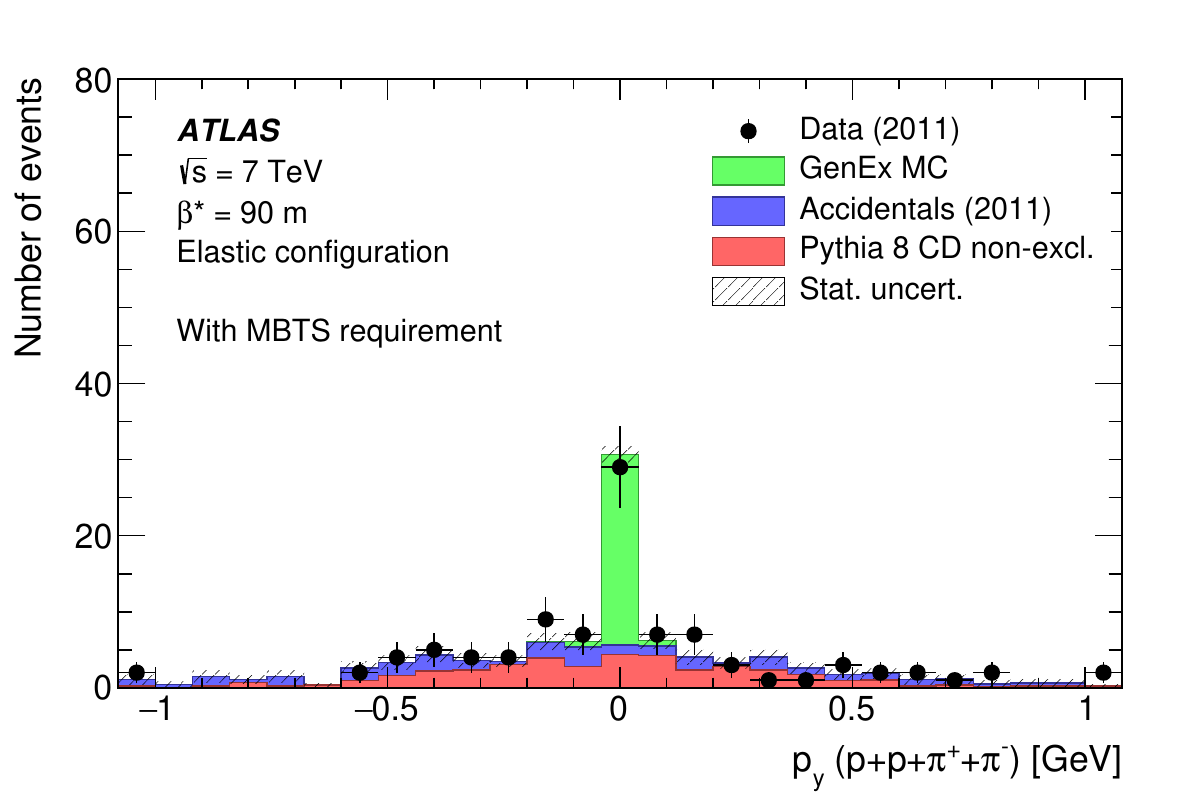}
\label{fig:background_after_MBTS}}
\caption{The summed momentum values in \(p_y\) after the pion and proton system selection: (a) before and (b) after the MBTS veto is applied (see  \Tab{\ref{tab:data_statistics}}). The histograms of the MC predictions and the accidentals are stacked. Statistical uncertainties of MC events and accidentals are combined. The underflow and overflow bins ($|p_y|>1$~\GeV) are shown.}
\label{fig:sum_py_w_wo_mbts}
\end{figure}

The very strong effect of the MBTS veto on background suppression is demonstrated also in \Fig{\ref{fig:sum_py_w_wo_mbts}}, which shows that the summed vertical ($y$-axis) track momenta exhibit a clear signal peak in the total $p_y$ of the event only after this veto is applied. \Fig{\ref{fig:background_before_MBTS}} includes events  with selections up to those on the ID and ALFA tracks listed in Table~\ref{tab:data_statistics},
while \Fig{\ref{fig:background_after_MBTS}} shows the same data and backgrounds but with the MBTS veto also applied.
The background was fitted outside the signal region, which approximately covers the central three plotted intervals.  The signal peak is now prominent.
 
\begin{figure}[ht!]
\centering
\subfloat[]{\includegraphics[width=0.49\textwidth]{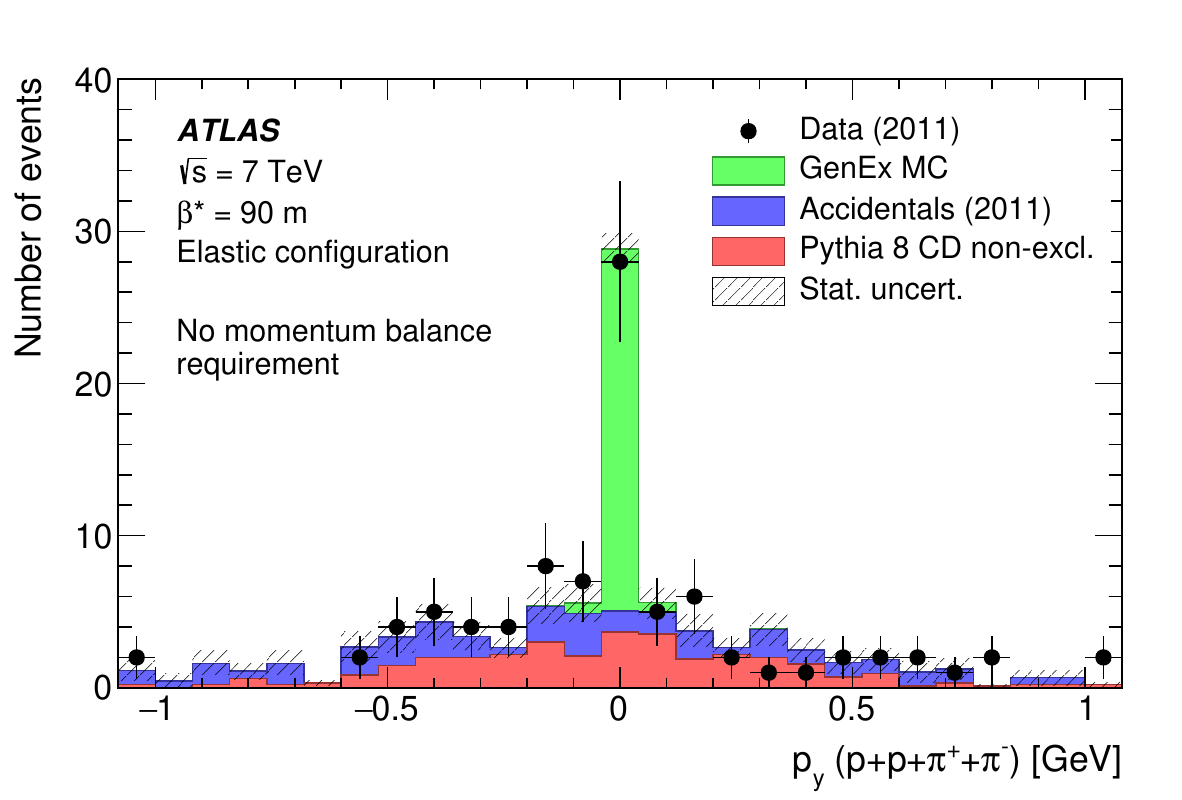}
\label{fig:TrkSumSyst_py_data_fit_switchOffPxPy_elastic}}
\subfloat[]{\includegraphics[width=0.49\textwidth]{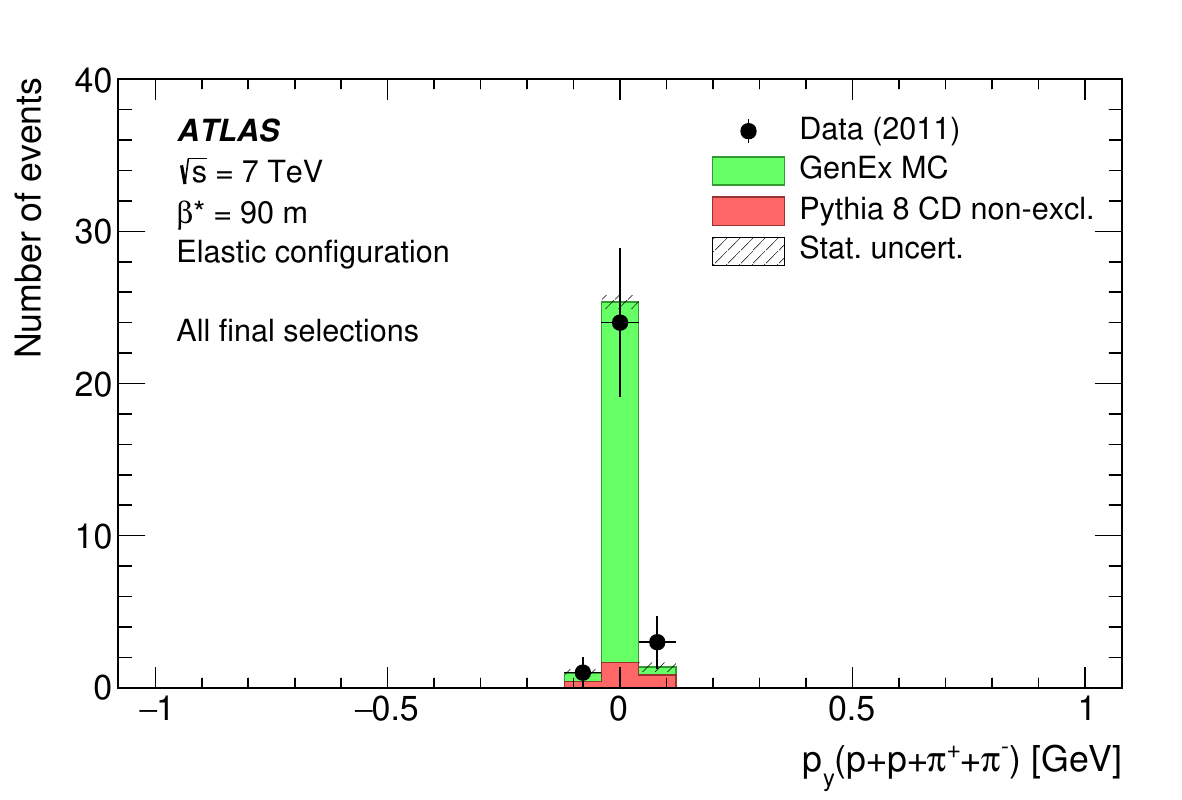}
\label{fig:TrkSumSyst_py_data_fit_final_elastic}}
\caption{Full final-state \(p_y\) momentum component of data, fitted to MC simulation and accidentals using \GENEX for the signal component and a background consisting of  accidentals and \PYTHIAe central diffraction (CD) without exclusive pion processes; in (a) all the final selections listed in \Tab{\ref{tab:cuts}} are applied except those on the momentum balance in the $x$ and $y$ projections, while in (b) all the final selections are applied. The histograms of the MC predictions and the accidentals are stacked. Statistical uncertainties of the MC events and accidentals are combined. The underflow and overflow bins ($|p_y|>1$~\GeV) are shown.}
\end{figure}

Since the data were taken during the special run with very low pile-up, the remaining backgrounds, after applying the event selections of \Sect{\ref{sec:data_analysis_event_selection}}, presumably originate from other central diffraction processes. For both the elastic and anti-elastic event configurations, the backgrounds due to these diffraction processes are estimated using events satisfying the complete set of selections listed in \Tab{\ref{tab:cuts}} except for the requirement on the momentum balance in the $x$ and $y$ projections. They are evaluated from a fit to the $p_y$ distribution of the data outside the signal region, using a model consisting of a background generated by \PYTHIAe  and accidentals.  This is shown in \Fig{\ref{fig:TrkSumSyst_py_data_fit_switchOffPxPy_elastic}}, while \Fig{\ref{fig:TrkSumSyst_py_data_fit_final_elastic}} shows the distribution with all the selections applied.
 
\subsection{Beam and ALFA modelling inefficiencies, correction procedure}
The implementation of the ALFA detector in the ATLAS simulation framework has some limitations.  The response can differ from that of the real detector, mainly due to imperfect implementation of signal and noise production mechanisms and the omission of electronic cross-talk between channels.
 
Since the simulation does not include the production of particle showers that are due to the interaction of forward-going particles with parts of the LHC (accelerator beam pipe, collimators, shielding of magnets, etc.), the \textit{uv}-condition
is introduced, effectively removing these showers from the data. This affects the signal, and the corresponding correction depends on the elastic or anti-elastic configuration (`conf') and the arm. A further correction is required for the clean track selection (Section \ref{sec:data_analysis_event_selection}).
 
If the MC description were perfect, the detector efficiency \(\varepsilon\), for a given arm and configuration, would be written as
\begin{equation}
\label{eq:topology_eff_mc}
\left(\varepsilon\right)^{\mathrm{conf}}_{\textrm{arm}} = \left[\left( \frac{N_{\textrm{trk}}}{N_{\textrm{acc}}}\right)^{\textrm{MC}}\right]_{\textrm{arm}}^{\mathrm{conf}}  \textrm{,}
\end{equation}
where $N_{\textrm{trk}}$ is the number of generated MC signal events with at least one track reconstructed in each of the detectors in a given configuration and ${N_{\textrm{acc}}}$ is the number of MC signal events within the acceptance of that configuration.
 
Since the MC simulation does not contain all relevant effects, the $uv$-condition and the clean track selection are introduced and Eq. (\ref{eq:topology_eff_mc}) is now rewritten as
\begin{equation}
\label{eq:topology_eff_mc_data}
\left(\varepsilon'\right)^{\mathrm{conf}}_{\textrm{arm}} = \left[ \left( \frac{N_{\textrm{trk}}}{N_{\textrm{uv}}}\right)^{\textrm{data}} \times \left( \frac{N_{\textrm{uv}}}{N_{\textrm{acc}}}\right)^{\textrm{MC}}\right]_{\textrm{arm}}^{\mathrm{conf}} \stackrel{\text{def}}{=} \left[r^{\textrm{data}} \times t^{\textrm{MC}} \right]_{\textrm{arm}}^{\mathrm{conf}} \textrm{,}
\end{equation}
where ${N_{\textrm{uv}}}$ is the number of events which pass the \textit{uv}-condition.
The first factor in Eq.~(\ref{eq:topology_eff_mc_data}) is given by physics that can be mismodelled by generators, and the second factor takes account of the clean track selection.
 
Denoting \(R_{\textrm{arm}}^{\mathrm{conf}} = \left(o^{\textrm{MC}}/o^{\textrm{data}}\right)_{\textrm{arm}}^{\mathrm{conf}}\), where the ratio \(o^{\textrm{X}} = N^{\textrm{X, clean}}/N^{\textrm{X, any}}\) gives the ratio of the number of events with clean tracks to all events in a given track sample X (MC or data), the combined correction factor \(f^{\mathrm{s+}\mathit{uv}\mathrm{, conf}}_{\textrm{arm}}\), where \(\mathrm{s+}\mathit{uv}\) denotes the combination of single tracks and $uv$-conditions, is written as
 
\begin{equation*}
f^{\mathrm{s+}\mathit{uv}\mathrm{, conf}}_{\textrm{arm}} = \left( \varepsilon / \varepsilon' \times R \right)_{\textrm{arm}}^{\mathrm{conf}}
\end{equation*}
 
which, using Eq.~(\ref{eq:topology_eff_mc_data}), can be re-written as
 
\begin{equation*}
f^{\mathrm{s+}\mathit{uv}\mathrm{, conf}}_{\textrm{arm}} = \left( r^{\textrm{MC}} / r ^{\textrm{data}} \times o^{\textrm{MC}}/o^{\textrm{data}} \right)_{\textrm{arm}}^{\textrm{conf}}	\textrm{.}
\end{equation*}
Its values are given in Table~\ref{tab:alfa_final_correction_single}.
 
The effects arising from detector efficiency and due to migrations across the fiducial phase-space boundary at particle and detector level are corrected for with a single  factor for each  configuration. This is adequate in view of the limited statistical precision of the data, given that the MC model describes the data well. The correction factor, \(f^{\mathrm{corr, conf}}_{\mathrm{arm}}\), is based on the \GENEX MC simulation, and for a given configuration it is defined as the ratio of the number of events, \(N_{\mathrm{detector\_level}}^{\mathrm{passed}}\), that pass the event selection criteria at the detector level, to the number of events \(N_{\mathrm{particle\_level}}^{\mathrm{passed}}\) that pass the fiducial region selection at the particle level, i.e.  	\(f^{\mathrm{corr, conf}}_{\mathrm{arm}} = \left({N_{\mathrm{detector\_level}}^{\mathrm{passed}}}/{N_{\mathrm{particle\_level}}^{\mathrm{passed}}}\right)^{\mathrm{conf}}_{\mathrm{arm}}\). The geometrical acceptance of each ALFA detector is $\approx$ 0.25, giving rise to the calculated combined values stated in \Tab{\ref{tab:alfa_final_correction_single}}.
 
\begin{minipage}{0.95\textwidth}
\centering
\captionof{table}{Correction factors for the single-track selection and \textit{uv}-condition, and the correction factor for MC simulation based on the \GENEX generator. Statistical uncertainties are quoted.}
\label{tab:alfa_final_correction_single}
\as{1.2}
\begin{tabular}{@{}lccc@{}}
\toprule
Configuration & arm & \(f^{\mathrm{s+}\mathit{uv}\mathrm{, conf}}_{\textrm{arm}}\) & \(f^{\mathrm{corr, conf}}_{\mathrm{arm}}\) \\
\midrule
\multirow{2}{5em}{Elastic} & 0 & \(1.097\pm0.015 \) & \(0.076\pm0.001 \) \\
& 1 & \(1.135\pm0.016 \) & \(0.070\pm0.001 \) \\
\multirow{2}{5em}{\mbox{Anti-elastic}} & 0 & \(1.229\pm0.115 \) & \(0.069\pm0.001 \) \\
& 1 & \(1.126\pm0.134 \) & \(0.068\pm0.001 \) \\
\bottomrule
\\
\end{tabular}
\end{minipage}

The visible cross-section for a given elastic or anti-elastic configuration is then determined as
 
\begin{equation*}
\sigma_{\piplus\piminus}^{\mathrm{vis,\ conf}} = \frac{1}{L} \times \left[ \left(N_{0}^{\mathrm{conf}}-B_{0}^{\mathrm{conf}}\right) \times C_{0}^{\mathrm{conf}} + \left(N_{1}^{\mathrm{conf}}-B_{1}^{\mathrm{conf}}\right) \times C_{1}^{\mathrm{conf}} \right] \times \frac{1}{\epsilon_{\mathrm{trig}}^{\mathrm{conf}} \cdot \epsilon_{\mathrm{DAQ}}}
\end{equation*}
 
where $L$ represents the integrated luminosity used; $N^{\mathrm{conf}}_{\mathrm{arm}}$ and $B^{\mathrm{conf}}_{\mathrm{arm}}$ are the  numbers of selected signal candidate events and calculated background events in each  arm; \( C^{\mathrm{conf}}_{\textrm{arm}} = f^{\mathrm{s+}\mathit{uv}\mathrm{, conf}}_{\textrm{arm}}/f^{\mathrm{corr, conf}}_{\mathrm{arm}}\);  $\epsilon_{\mathrm{trig}}^{\mathrm{conf}}$ is the trigger efficiency for the given configuration; $\epsilon_{\mathrm{DAQ}}$ is the dead-time correction factor, as defined and determined in Ref.~\cite{Aad:2014dca}.
 
\subsection{Kinematic distributions}
\begin{figure}[ht!]
\centering
\includegraphics[width=0.9\textwidth]{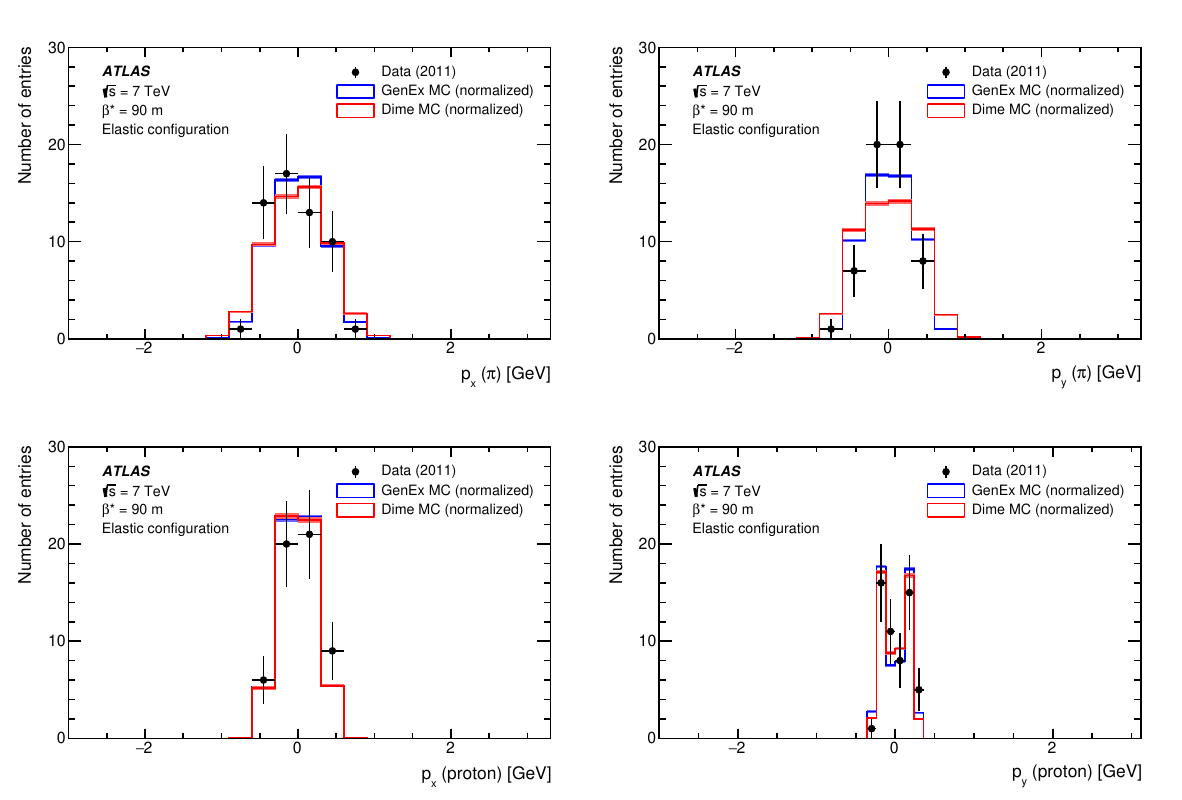}
\caption{Distributions, for the selected events, of \(p_x\) and \(p_y\) components of the pion tracks reconstructed in the ATLAS central detector (upper row), and \(p_x\) and \(p_y\) components of the  proton tracks reconstructed in the ALFA detectors (lower row) for the data, \GENEX MC and \DIME MC simulations after applying all the event selections, but without a background subtraction applied to the data (\(\approx 10\%\)). Each of the MC samples is scaled to the data. Statistical uncertainties are shown only for the data; they are negligible for the MC predictions.}
\label{fig:CombinedPlots_oppoSign_DPEnonExcl_lattice_KinReg}
\end{figure}

\begin{figure}
\centering
\subfloat[]{\includegraphics[width=0.45\textwidth]{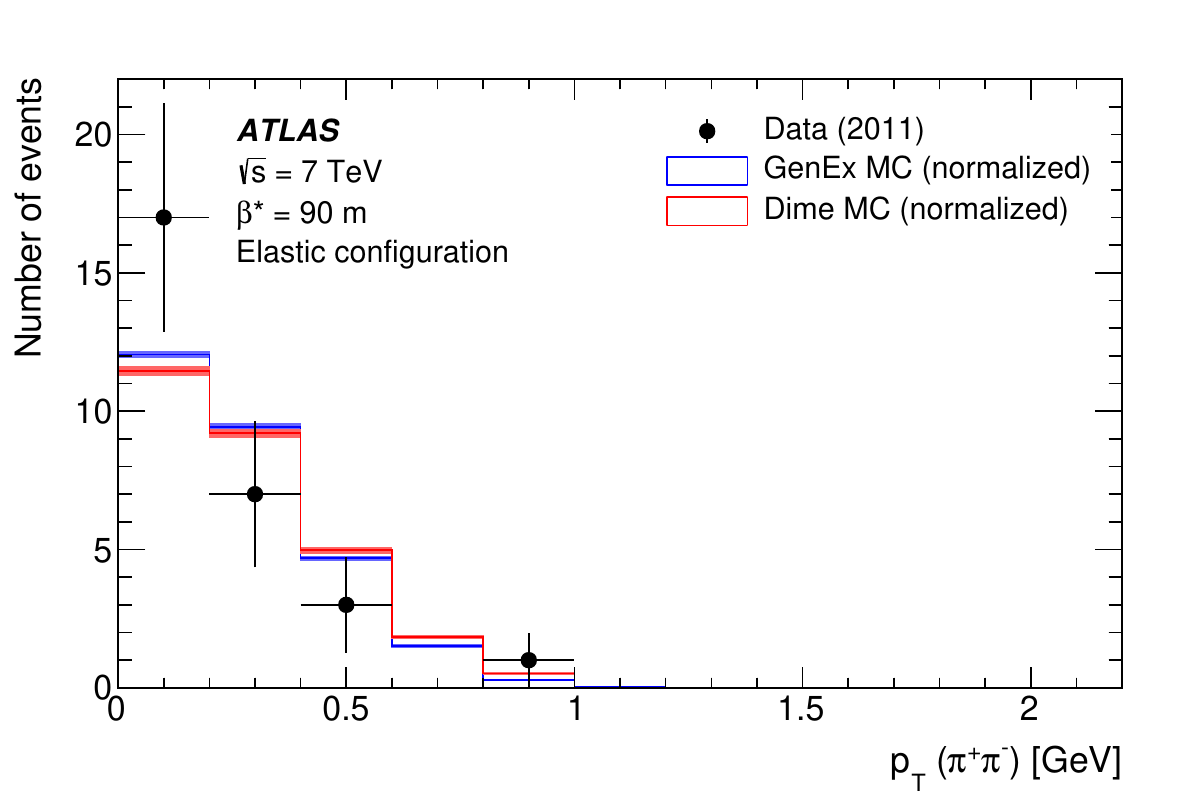}
}
\subfloat[]{\includegraphics[width=0.45\textwidth]{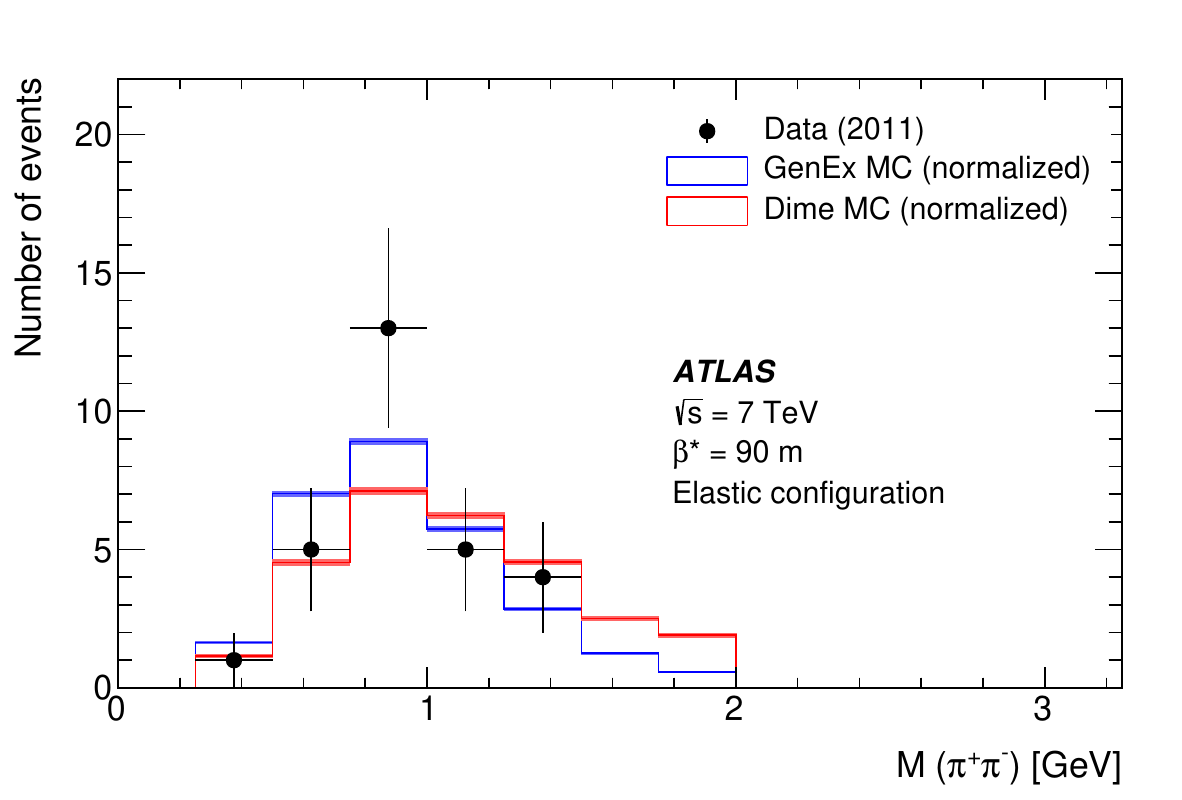}
}
\caption{Distributions of (a) pion-pair transverse momentum and (b) pion-pair invariant mass for the data, \GENEX MC and \DIME MC simulations after applying all event selections, but without a background subtraction applied to the data. Each of the MC samples is normalized to the data. Statistical uncertainties are shown only for the data; they are negligible for the MC predictions.}
\label{fig:CombinedPlots_oppoSign_DPEnonExcl_lattice_KinReg_pt_inv_mass}
\end{figure}

Taking into account the precision of the reconstruction,
the kinematic distributions of the pions and protons are well modelled by the MC simulations. \Figrange{\ref{fig:CombinedPlots_oppoSign_DPEnonExcl_lattice_KinReg}}{\ref{fig:CombinedPlots_oppoSign_DPEnonExcl_lattice_KinReg_pt_inv_mass}} show comparisons
of the distributions of the momentum components of the reconstructed pions and protons, and the pion-pair rapidity, acoplanarity, transverse momentum and invariant mass,  for the data and the \GENEX MC and \DIME MC simulations.
The distributions are presented after applying all the event selections, but without applying a background subtraction (\(\approx 10\%\)) to the data. In view of the low number of events for the anti-elastic configuration, only distributions for the elastic configuration are shown.
 
\begin{figure}
\centering
\subfloat[]{\includegraphics[width=0.45\textwidth]{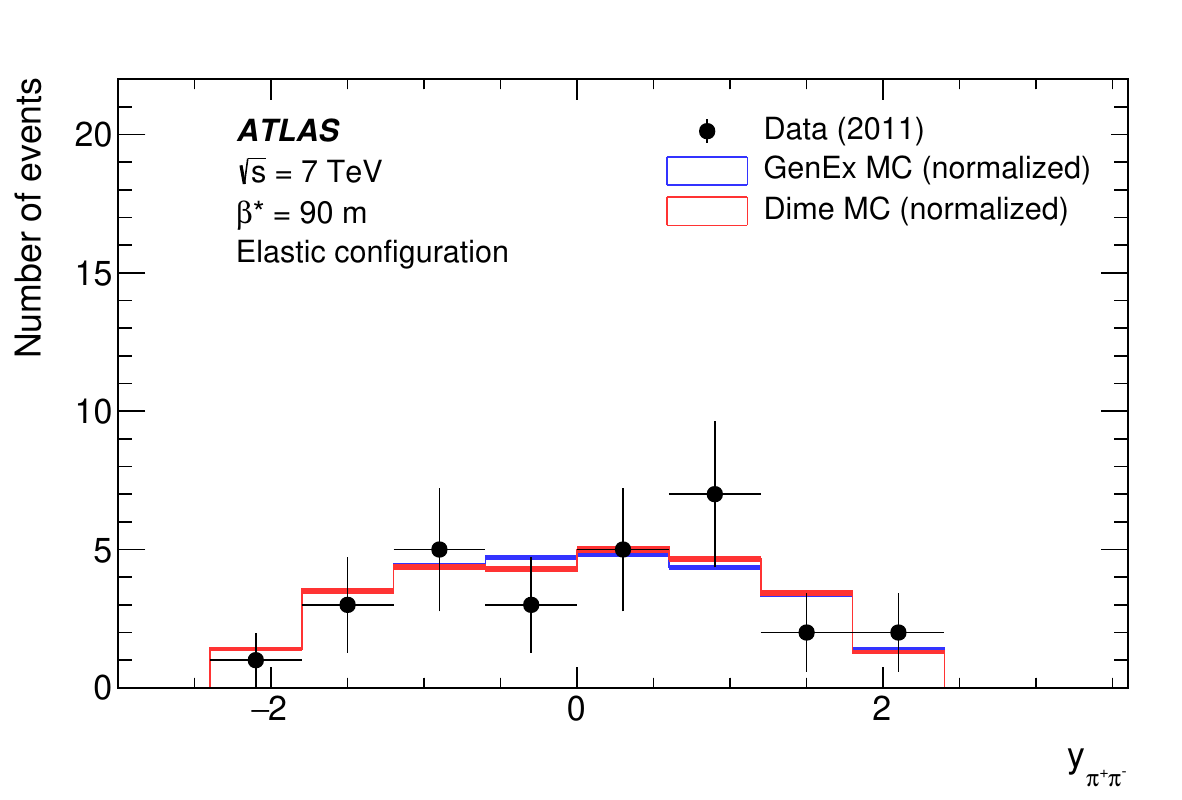}
}
\subfloat[]{\includegraphics[width=0.45\textwidth]{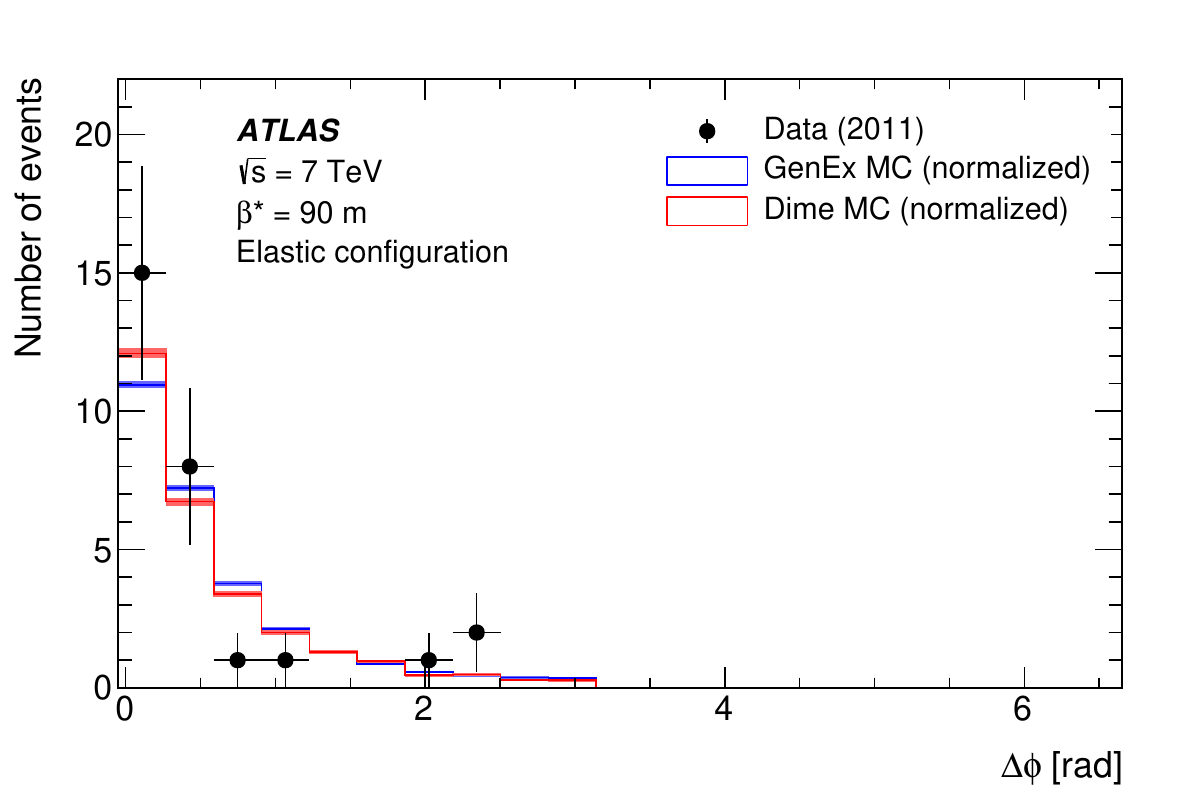}
}
\caption{Distributions of (a) pion-pair rapidity and (b) pion-pair acoplanarity for the data, \GENEX MC and \DIME MC simulations after applying all event selections, but without a background subtraction applied to the data. Each of the MC samples is normalized to the data. Statistical uncertainties are shown only for the data; they are negligible for the MC predictions.}
\end{figure}
 
\subsection{Systematic uncertainties}
 
The overall systematic uncertainty on the exclusive pion-pair cross-section measurement is estimated as
\(\uncertupdown{6.4}{4.2}\)\,\(\%\) for the elastic configuration and
\(\uncertupdown{6.0}{4.4}\)\,\(\%\)
for the anti-elastic configuration.  \Tab{\ref{tab:syst_uncert_summary}} summarizes the main sources of systematic uncertainty. The difference between the relative uncertainties of the two configurations comes mainly from data-driven systematic uncertainties.  Some of the MC studies made use of slightly enlarged fiducial region criteria to estimate systematic uncertainties.
 
\textbf{Background determination.} The systematic uncertainty is determined as half of the difference between the background estimates for the standard signal region, \(3.5\sigma_{x,y}\) around zero in $p_x$ and $p_y$, and for the case when the signal region is increased to \(5\sigma_{x,y}\). It is symmetrized.
 
\textbf{LHC beam energy.} The systematic uncertainty due to that of the nominal beam energy is based on a beam energy variation of about \(\pm0.1\%\)~\cite{Wenninger:1546734}.
 
\textbf{Thicknesses of material.} An uncertainty in the amount of material in the ID is calculated by changing this  by \(+5\%\) and \(+10\%\) in the whole ID and by \(+10\%\) in the pixel detector to determine the difference between the lowest and the highest value of the corrected exclusive pion-pair cross-section \cite{STDM-2010-06}.
 
\textbf{MBTS Veto.} The MBTS veto uncertainty determination is based on the corresponding procedure used in the measurement of the inelastic proton--proton cross-section at \(\sqrt s\) = \SI{7}{\TeV} \cite{STDM-2010-11}.
 
\textbf{Selection criteria.} The uncertainty in the efficiency due to the selection criteria is determined from the fraction of lost signal. The overall uncertainty in the ALFA reconstruction efficiency is determined by varying the event selection criteria and summing the corresponding individual uncertainties in quadrature.
 
\textbf{Correction factors.} An uncertainty in the values of $f^{\mathrm{s+}\mathit{uv}\mathrm{, conf}}_{\textrm{arm}}$ arises from the restricted selection of data that is used to evaluate this factor in each arm.  Although of statistical origin, this is treated as a systematic uncertainty.
 
\textbf{Beam optics.} This analysis uses data at a collision energy of \SI{7}{\TeV} and with high $\betastar = \SI{90}{m}$ beam optics, referred to as the nominal optics. Based on data measurements, an adjustment is made to the current-to-field calibration offset between the Q1Q3 and Q2 quadrupole magnets to derive the so-called effective optics \cite{Aad:2014dca}. Studies related to both optics are performed: the effective and the nominal optics are used in MC simulation to calculate all the corrections applied in reconstructing the data. The nominal optics is used as the baseline; the difference in the cross-section values is used to determine the systematic uncertainties.
 
\textbf{Modelling.} The analysis of the data uses the generator \GENEX as the baseline for modelling the signal sample. The full analysis is repeated using the generator \DIME, and the difference between the resulting cross-sections is taken as an uncertainty due to the modelling and is applied symmetrically.
 
\begin{minipage}{0.95\textwidth}
\centering
\captionof{table}{Summary of the exclusive pion-pair cross-section systematic uncertainties.}
\label{tab:syst_uncert_summary}
\as{1.2}
\begin{tabular}{@{}lrr@{}}
\toprule
& \multicolumn{2}{c}{\bfseries Uncertainty [\%]} \\
\centercell{\bfseries Source of uncertainty} & \centercell{\bfseries elastic} & \centercell{\bfseries anti-elastic} \\
\cline{1-3}
\midrule
Trigger efficiency \(\epsilon_{\mathrm{trig}}\) & \textpm0.1 & \textpm0.3 \\
Background determination & \textpm3.5 & \textpm3.5 \\
 
Signal and background corrections:   & {} & {} \\
\ \ \ \ Beam energy & \textpm0.1 & \textpm0.1 \\
\ \ \ \ ID material & +4.8 & +4.1\\
\ \ \ \ Veto on MBTS signal & \textpm1.3 & \textpm2.0 \\
\ \ \ \ ALFA single-track selection & \textpm0.9 & \textpm0.9 \\
\ \ \ \ ALFA reconstruction efficiency & \textpm0.9 & \textpm0.8 \\
\ \ \ \ ALFA geometry selection & \textpm0.5 & \textpm0.5 \\
\ \ \ \ Optics & \textpm1.1 & \textpm1.0 \\
\midrule
Overall systematic uncertainty & {\vspace{3pt} \Large{\(\uncertupdown{6.4}{4.2}\)}}  & { \Large{\(\uncertupdown{6.0}{4.4}\)}} \\
\cline{1-3}
Statistical uncertainty & \textpm21.2 & \textpm61.6 \\
\cline{1-3}
Theoretical modelling & \textpm2.8 & \textpm8.0 \\
\cline{1-3}
Luminosity  & \textpm1.2 & \textpm1.2 \\
\bottomrule
\end{tabular}
\end{minipage}

\section{Results and discussion}
 
The cross-section for exclusive pion pair production is measured for two non-overlapping configurations, elastic and anti-elastic. The fiducial region is defined in Section~\ref{sec:data_analysis_event_selection_description} and the results are presented in \Tab{\ref{tab:cross_section_comp}}.

\begin{minipage}{0.95\textwidth}
\centering
\captionof{table}{Exclusive pion-pair cross-section within the defined fiducial region for elastic and anti-elastic configurations, for the measurement and MC simulations. The statistical uncertainties of the MC simulations are negligible. The \GENEX predictions have been multiplied by an absorptive correction
factor of 0.22 suggested by the author of the generator.
}
\label{tab:cross_section_comp}
\as{1.3}
\begin{tabular}{ll}
\toprule
\multicolumn{2}{c}{\bfseries Exclusive \(\piplus\piminus\) cross-section \([\SI{}{\micro\mathrm{b}}]\)} \\
\midrule{\small{\textbf{Elastic configuration}}} & \\
{\small Measurement } & { \small $ 4.8 \pm 1.0 \mathrm{\ (stat)} \uncertupdown{0.3}{0.2} \mathrm{\ (syst)} \pm 0.1 \mathrm{\ (lumi)} \pm 0.1 \mathrm{\ (model)} $ } \\
{\small \GENEX  $ \times \ 0.22 \ \textrm{ (\small{absorptive correction})}$} & { \small $ 1.5 $ } \\
{\small \DIME } & { \small $1.6 $} \\
{\small \textbf{Anti-elastic configuration} } &  \\
{\small measurement } & { \small $9 \pm 6 \mathrm{\ (stat)} \uncertupdown{1}{1} \mathrm{\ (syst)} \pm 1 \mathrm{\ (lumi)} \pm 1 \mathrm{\ (model)} $ } \\
{\small \GENEX $ \times \ 0.22 \textrm{\ (\small{absorptive correction})} $ } & { \small $ 2 $} \\
{\small \DIME} & { \small $3 $} \\
\bottomrule
\end{tabular}
\end{minipage}

There are 28 candidate signal events in the elastic configuration and 3 in the anti-elastic configuration (see Table~\ref{tab:data_statistics}). These include an estimated
2.9 background events in the elastic configuration ($\approx$ 10\%) and 0.2 background events ($\approx$ 6\%) in the anti-elastic configuration.
 
The results may be compared with  cross-sections provided by MC generators using the same fiducial region definition:
\(\SI{1.5}{\micro\mathrm{b}}\) (elastic configuration) and
\(\SI{2}{\micro\mathrm{b}}\) (anti-elastic configuration) in the case of \GENEX (these have been multiplied by an absorptive correction, abs.\ corr.),  and
\(\SI{1.6}{\micro\mathrm{b}}\) (elastic configuration) and
\(\SI{3}{\micro\mathrm{b}}\) (anti-elastic configuration) in the case of  \DIME.
The absorptive correction~\cite{Harland-Lang:2013dia} applied here to \GENEX\ covers possible rescatterings of the particles entering the process and complicates the simple picture in Figure~\ref{fig:pipi_schematics}. The value used of 0.22  was communicated by its author; the correction is not included in \GENEX\ itself and is, in general, kinematics and model dependent. It is not used in the experimental analysis, since the MC distributions are normalized to data, but only in the present comparison with the results.
The measured anti-elastic cross-section has a  large statistical uncertainty, and is not statistically different from zero, which prevents strong conclusions from being drawn.
 
\section{Conclusion}
 
A first measurement of the purely exclusive pion-pair cross-section at the LHC is presented, using $\SI{80}{\micro\mathrm{b}}^{-1}$ of \(\rts=\SI{7}{\TeV}\) low-luminosity $pp$ collision data collected by the ATLAS detector.  The limited statistical precision obtained  excludes the present possibility of tuning or excluding any of the existing physical models for the process, but the two models that are employed here, \GENEX and \DIME, evaluated with the parameters provided by their authors,
provide preliminary theoretical estimates.
The use of the forward ALFA detectors distinguishes the present measurement from others performed at comparable energies, since both outgoing protons have been measured directly,
and used to suppress contamination from beam remnants. This measurement demonstrates the potential to measure exclusive diffractive hadronic processes using forward sub-detectors in combination with the ATLAS central detector.

\section*{Acknowledgements}

% The next lines are included from the .//acknowledgements/Acknowledgements.tex input file

We thank CERN for the very successful operation of the LHC, as well as the
support staff from our institutions without whom ATLAS could not be
operated efficiently.
 
We acknowledge the support of
ANPCyT, Argentina;
YerPhI, Armenia;
ARC, Australia;
BMWFW and FWF, Austria;
ANAS, Azerbaijan;
CNPq and FAPESP, Brazil;
NSERC, NRC and CFI, Canada;
CERN;
ANID, Chile;
CAS, MOST and NSFC, China;
Minciencias, Colombia;
MEYS CR, Czech Republic;
DNRF and DNSRC, Denmark;
IN2P3-CNRS and CEA-DRF/IRFU, France;
SRNSFG, Georgia;
BMBF, HGF and MPG, Germany;
GSRI, Greece;
RGC and Hong Kong SAR, China;
ISF and Benoziyo Center, Israel;
INFN, Italy;
MEXT and JSPS, Japan;
CNRST, Morocco;
NWO, Netherlands;
RCN, Norway;
MEiN, Poland;
FCT, Portugal;
MNE/IFA, Romania;
MESTD, Serbia;
MSSR, Slovakia;
ARRS and MIZ\v{S}, Slovenia;
DSI/NRF, South Africa;
MICINN, Spain;
SRC and Wallenberg Foundation, Sweden;
SERI, SNSF and Cantons of Bern and Geneva, Switzerland;
MOST, Taiwan;
TENMAK, T\"urkiye;
STFC, United Kingdom;
DOE and NSF, United States of America.
In addition, individual groups and members have received support from
BCKDF, CANARIE, Compute Canada and CRC, Canada;
PRIMUS 21/SCI/017 and UNCE SCI/013, Czech Republic;
COST, ERC, ERDF, Horizon 2020 and Marie Sk{\l}odowska-Curie Actions, European Union;
Investissements d'Avenir Labex, Investissements d'Avenir Idex and ANR, France;
DFG and AvH Foundation, Germany;
Herakleitos, Thales and Aristeia programmes co-financed by EU-ESF and the Greek NSRF, Greece;
BSF-NSF and MINERVA, Israel;
Norwegian Financial Mechanism 2014-2021, Norway;
NCN and NAWA, Poland;
La Caixa Banking Foundation, CERCA Programme Generalitat de Catalunya and PROMETEO and GenT Programmes Generalitat Valenciana, Spain;
G\"{o}ran Gustafssons Stiftelse, Sweden;
The Royal Society and Leverhulme Trust, United Kingdom.
 
The crucial computing support from all WLCG partners is acknowledged gratefully, in particular from CERN, the ATLAS Tier-1 facilities at TRIUMF (Canada), NDGF (Denmark, Norway, Sweden), CC-IN2P3 (France), KIT/GridKA (Germany), INFN-CNAF (Italy), NL-T1 (Netherlands), PIC (Spain), ASGC (Taiwan), RAL (UK) and BNL (USA), the Tier-2 facilities worldwide and large non-WLCG resource providers. Major contributors of computing resources are listed in Ref.~\cite{ATL-SOFT-PUB-2023-001}.

% End of text imported from the .//acknowledgements/Acknowledgements.tex input file

\printbibliography
 
\clearpage
 
% ATLAS Collaboration author list
% Reference date of STDM-2017-07 is 2022-05-23
% Author list last updated on date 17-JAN-24
% Data extracted on 17-Jan-2024 for paper reference STDM-2017-07
% at 4:27pm
 
\begin{flushleft}
\hypersetup{urlcolor=black}
{\Large The ATLAS Collaboration}

\bigskip

\AtlasOrcid[0000-0002-6665-4934]{G.~Aad}$^\textrm{\scriptsize 101}$,
\AtlasOrcid[0000-0002-5888-2734]{B.~Abbott}$^\textrm{\scriptsize 119}$,
\AtlasOrcid[0000-0002-7248-3203]{D.C.~Abbott}$^\textrm{\scriptsize 102}$,
\AtlasOrcid[0000-0002-1002-1652]{K.~Abeling}$^\textrm{\scriptsize 55}$,
\AtlasOrcid[0000-0002-8496-9294]{S.H.~Abidi}$^\textrm{\scriptsize 29}$,
\AtlasOrcid[0000-0002-9987-2292]{A.~Aboulhorma}$^\textrm{\scriptsize 35e}$,
\AtlasOrcid[0000-0001-5329-6640]{H.~Abramowicz}$^\textrm{\scriptsize 150}$,
\AtlasOrcid[0000-0002-1599-2896]{H.~Abreu}$^\textrm{\scriptsize 149}$,
\AtlasOrcid[0000-0003-0403-3697]{Y.~Abulaiti}$^\textrm{\scriptsize 116}$,
\AtlasOrcid[0000-0003-0762-7204]{A.C.~Abusleme~Hoffman}$^\textrm{\scriptsize 136a}$
% ...
\bigskip
\\

% ...
$^{29}$Physics Department, Brookhaven National Laboratory, Upton NY; United States of America.\\
% ...
$^{35}$$^{(a)}$Facult\'e des Sciences Ain Chock, R\'eseau Universitaire de Physique des Hautes Energies - Universit\'e Hassan II, Casablanca;$^{(b)}$Facult\'{e} des Sciences, Universit\'{e} Ibn-Tofail, K\'{e}nitra;$^{(c)}$Facult\'e des Sciences Semlalia, Universit\'e Cadi Ayyad, LPHEA-Marrakech;$^{(d)}$LPMR, Facult\'e des Sciences, Universit\'e Mohamed Premier, Oujda;$^{(e)}$Facult\'e des sciences, Universit\'e Mohammed V, Rabat;$^{(f)}$Institute of Applied Physics, Mohammed VI Polytechnic University, Ben Guerir; Morocco.\\
% ...
$^{55}$II. Physikalisches Institut, Georg-August-Universit\"{a}t G\"ottingen, G\"ottingen; Germany.\\
% ...
$^{101}$CPPM, Aix-Marseille Universit\'e, CNRS/IN2P3, Marseille; France.\\
$^{102}$Department of Physics, University of Massachusetts, Amherst MA; United States of America.\\
% ...
$^{116}$Department of Physics, New York University, New York NY; United States of America.\\
% ...
$^{119}$Homer L. Dodge Department of Physics and Astronomy, University of Oklahoma, Norman OK; United States of America.\\
% ...
$^{136}$$^{(a)}$Departamento de F\'isica, Pontificia Universidad Cat\'olica de Chile, Santiago;$^{(b)}$Millennium Institute for Subatomic physics at high energy frontier (SAPHIR), Santiago;$^{(c)}$Instituto de Investigaci\'on Multidisciplinario en Ciencia y Tecnolog\'ia, y Departamento de F\'isica, Universidad de La Serena;$^{(d)}$Universidad Andres Bello, Department of Physics, Santiago;$^{(e)}$Instituto de Alta Investigaci\'on, Universidad de Tarapac\'a, Arica;$^{(f)}$Departamento de F\'isica, Universidad T\'ecnica Federico Santa Mar\'ia, Valpara\'iso; Chile.\\
% ...
$^{149}$Department of Physics, Technion, Israel Institute of Technology, Haifa; Israel.\\
$^{150}$Raymond and Beverly Sackler School of Physics and Astronomy, Tel Aviv University, Tel Aviv; Israel.\\
% ...
$^{a}$ Also Affiliated with an institute covered by a cooperation agreement with CERN.\\
$^{b}$ Also at An-Najah National University, Nablus; Palestine.\\
$^{c}$ Also at Borough of Manhattan Community College, City University of New York, New York NY; United States of America.\\
$^{d}$ Also at Bruno Kessler Foundation, Trento; Italy.\\
$^{e}$ Also at Center for High Energy Physics, Peking University; China.\\
$^{f}$ Also at Centro Studi e Ricerche Enrico Fermi; Italy.\\
% ...

\end{flushleft}

% Created with Glance <Atlas.Glance@cern.ch>

\end{document}